\documentstyle[12pt]{article}
\newcommand{\be}{\begin{equation}}
\newcommand{\ee}{\end{equation}}
\newcommand{\bea}{\begin{eqnarray}}
\newcommand{\eea}{\end{eqnarray}}

\newcommand{\eps}{\varepsilon}

\newcommand{\half}{{\scriptstyle{{1\over 2}}}}

\newcommand{\quart}{{\scriptstyle{{1\over 4}}}}

\newcommand{\real}{\relax{\rm I\kern-.18em R}}
\newcommand{\Tr}{{\rm Tr}}
\newcommand{\tr}{{\rm tr}}
\newcommand{\mod}{{\rm mod}}
\newcommand{\acosh}{{\rm acosh}}

\def\myre{{\rm Re}}
\def\ba{\begin{array}}
\def\ea{\end{array}}

\begin{document}
\vskip-1cm
\hfill FTUAM-99-6; IFT-UAM/CSIC-99-8
\vskip0mm
\hfill INLO-PUB-6/99
\vskip5mm
\begin{center}
{\LARGE{\bf{\underline{Calorons on the lattice - a new perspective}}}}\\
\vspace*{7mm}{\large
Margarita Garc\'{\i}a P\'erez$^{(a)}$, Antonio Gonz\'alez-Arroyo$^{(a,b)}$,\\
Alvaro Montero$^{(a)}$ and Pierre van Baal$^{(c)}$\\}
\vspace*{5mm}
($a$) Departamento de F\'{\i}sica Te\'{o}rica C-XI,\\
Universidad Aut\'{o}noma de Madrid,\\
28049 Madrid, Spain.\\
\vspace*{3mm}
($b$) Instituto de F\'{\i}sica Te\'{o}rica C-XVI,\\
Universidad Aut\'{o}noma de Madrid,\\
28049 Madrid, Spain.\\
\vspace*{3mm}
($c$) Instituut-Lorentz for Theoretical Physics,\\
University of Leiden, PO Box 9506,\\
NL-2300 RA Leiden, The Netherlands.
\end{center}
\vspace*{2mm}{\narrower{\noindent
\underline{Abstract:} We discuss the manifestation of instanton and monopole
solutions on a periodic lattice at finite temperature and their relation to
the infinite volume analytic caloron solutions with asymptotic non-trivial 
Polyakov loops. As a tool we use improved cooling and twisted boundary 
conditions. Typically we find $2Q$ lumps for topological charge $Q$. These
lumps are BPS monopoles.}\par}

\section{Introduction}

Calorons are characterised by their holonomy, defined by the value of the 
Polyakov loop at spatial infinity. When non-trivial, it resolves the fact that 
a caloron is built from constituent monopoles, their mass ratios directly 
determined by the holonomy~\cite{KSu2,LeeL}. These solutions differ from the 
(deformed) instantons described by the Harrington-Shepard solution~\cite{HaSh},
for which the holonomy is trivial. What we find by (improved~\cite{UsCo}) 
cooling on a finite lattice, to relatively high accuracy, is $SU(2)$ 
configurations that fit these infinite volume caloron solutions for arbitrary 
constituent monopole mass ratios. Twist~\cite{THT4} in the time direction 
constrains the masses of the two constituent monopoles to be equal. 

The constituent nature becomes evident when the instanton scale parameter $\rho$
is larger than the time extent $\beta$ (inverse temperature) of the system. The
masses of the monopoles are for $SU(2)$ proportional~\cite{KSu2} to $\omega$
and $\half-\omega$, where $\omega$ ($0\leq\omega\leq\half$) follows from the 
trace of the 
holonomy: $2\cos(2\pi\omega)$. The distance between the monopole constituents 
is given by $\pi\rho^2/\beta$. At $\rho/\beta\ll 1$ the constituents therefore hide 
deep inside the core of the instanton and the non-trivial holonomy plays no 
discernible role. But for $\rho/\beta\gg 1$ the situation is opposite; the 
instanton becomes static and will dissolve in two BPS monopoles~\cite{THPo,BPS}.
The transition occurs~\cite{KrBo,Brow} for $\half \beta<\rho<\beta$. When, however, the 
holonomy is trivial one of the monopoles is massless and will hide in the 
background. 

Charge one $SU(N)$ calorons have $N$ constituent monopoles~\cite{KSuN} for
non-trivial holonomy. These have the same location in time, but the spatial
position of each constituent monopole can be arbitrary. There are (at fixed
holonomy) $N-1$ phases associated to the residual $U(1)^{N-1}$ gauge symmetry
that leaves the holonomy invariant. The total number of parameters describing
these calorons is therefore $4N$. One may speculate that the $N-1$ phases
are replaced in a finite volume by the holonomy itself, indeed described by
$N-1$ eigenvalues taking values in $U(1)$ ($\exp(2\pi i\omega)$ for $SU(2)$).
Also it is likely that, in general, a charge $Q$ caloron  is characterised by
$NQ$ constituent monopoles, which we confirm for a $Q=2$ caloron solution
obtained from cooling. At zero temperature it is tempting to explain the
$4NQ$ parameters of an $SU(N)$ charge $Q$ instanton in terms of the positions
of $NQ$ objects~\cite{Ton1}. Indeed, there are charge $Q=1/N$ instanton
solutions on a torus with twisted boundary conditions, whose four parameters
specify its position~\cite{MaTo}. Subdividing a given finite volume in boxes
with the appropriate twisted boundary conditions, such that each cell supports
a $Q=1/N$ instanton, provides an exact solution that has $NQ$ lumps. In
ref.~\cite{Ton1} it is suggested that  a {\em typical}  self-dual configuration
would appear as an ensemble of $N$ randomly placed  lumps of charge $Q=1/N$,
whose locations would account  for the $4NQ$ parameters. The results at finite
temperature presented here suggest that the assignment of $Q=1/N$ charge 
to  each lump might only hold on the average.

Our results point to the usefulness of studying the dynamical role of these 
configurations. A first attempt in that direction is hampered by the fact that 
at high temperatures where the constituent monopoles should be well separated, 
the fluctuations are so large that on average topological charge cannot be 
supported over large enough domains of space-time to capture the configurations
with cooling~\cite{MaNuPh}. 

On a semiclassical basis one is tempted to argue against non-trivial holonomy. 
It polarises the vacuum at infinity and raises the energy density above the 
one with a trivial holonomy~\cite{GrPiYa}. But now we have seen that these BPS 
bound states can be supported in a finite volume, it is time to acknowledge 
this as an irrelevant objection, given the non-perturbative and non-trivial 
nature of the QCD vacuum. As a consequence, constituent monopoles, at least at 
high temperatures, are tangible objects that do not depend on a choice of 
Abelian projection~\cite{THAbPr}, which till now has been used to address the 
monopole content of the theory~\cite{AbPrRev}. In extracting the non-trivial 
topological content of the theory constituent monopoles introduce an extra 
parameter: their mass, $16\pi^2\omega/\beta$. Up to now only the {\em maximal} 
mass, $8\pi^2/\beta$, of such a BPS monopole was considered. It arises in terms 
of the caloron with trivial holonomy, described by the Harrington-Shepard 
solution~\cite{HaSh}. Rossi~\cite{Rossi} showed that at high temperature, 
equivalent to a large scale parameter, this solution indeed becomes a BPS 
monopole~\cite{BPS}.

In section 2 we discuss the numerical procedure of constructing the 
configurations. Apart from cooling with improved actions, twisted boundary 
conditions are used as a tool for biasing the cooling towards non-trivial 
holonomy. The twist can then be removed, while preserving the non-trivial 
holonomy and the constituent monopole nature of the configuration, although 
it should be pointed out that no exact charge one instanton solutions can
exist on $T^4$, which {\em remains} true at finite temperature. Interesting 
in this respect is that the well-established $Q=1/2$ instanton solutions that 
occur with suitable combinations of spatial and temporal twists (so-called 
non-orthogonal twist), can be argued to become a {\em single} static BPS 
monopole in the infinite volume limit at finite temperature. This is discussed 
in section 3. Configurations of higher charge are discussed in section 4 and 
we conclude with some speculations and possible applications. An appendix 
summarises the formulae for the $SU(2)$ analytic caloron solutions.

\section{Non-trivial holonomy from time-twist}

For finite temperature ($T=1/\beta$) and volume ($L^3$, $L\gg \beta$) caloron 
configurations with non-trivial holonomy were discovered  on lattices with
twisted boundary conditions. Starting with a random configuration and after
applying  a standard cooling algorithm one frequently reaches $Q=1$ self-dual
configurations which are stable under many cooling steps.  These configurations
are later analysed. An automatic peak-searching routine identified one or 
(actually more frequently) two lumps in them. These are our candidate caloron
configurations on the lattice. Below, we will show how twist in the time direction
can help in bringing about non-trivial holonomy.

To appreciate the ease with which twist can be implemented on the lattice, and 
because twist has been a very useful tool~\cite{UsCo,TonRev,Buck}, neglected 
by large parts of the lattice community, we think it is useful to review the 
notion of twist-carrying plaquettes that introduce twist by modifying the 
lattice action~\cite{GrJuKA}, but not the measure. In the initial formulation 
of 't Hooft~\cite{THT4}, $SU(N)$ twisted boundary conditions were implemented 
by defining gauge functions $\Omega_\mu(x)$ (which are assumed independent of 
$x_\mu$), such that with $a^\mu$ the periods of the torus in the four 
directions ($a_\mu^\nu=L_\mu\delta_{\mu\nu}$)
\be
U_\nu(x+a^\mu)= \Omega_\mu(x)\, U_\nu(x)\, \Omega^\dagger_\mu(x+\hat\nu),
\ee
here re-formulated for a lattice of size $\prod_\mu N_\mu$. Calculating
$U_\nu(x+a^\mu+a^\lambda)$ in two ways shows that for all $x$ one should have
\be
\Omega_\mu(x+a^\lambda)\, \Omega_\lambda(x)=Z_{\mu\lambda}\,
\Omega_\lambda(x+a^\mu)\,
\Omega_\mu(x),
\ee
with $Z_{\mu\lambda}=\exp(2\pi i n_{\mu\lambda}/N)$ an element of the center 
of the gauge group. (We define $k_i=n_{0i}$ and $m_i=\half\varepsilon_{ijk}
n_{jk}$ to distinguish the twist in the time and space directions respectively).
The center freedom arises because $U_\mu(x)$ is invariant under constant center 
gauge transformations (i.e. the gauge field is in the adjoint representation). 
In the presence of site variables (fields in the fundamental representation) 
one is required to put all $Z_{\mu\nu}$ equal to 1.
We now perform the following change of variables~\cite{GrJuKA}
\be
U^\prime_\mu(x)=U_\mu(x)\, \Omega_\mu(x),\quad \mbox{for} \quad x_\mu=N_\mu-1.
\ee
As a consequence, the plaquettes at $x_\lambda\!=\!N_\lambda-1$ and $x_\mu\!=\!N_\mu-1$ 
(for any value of the other two components of $x$) can be shown to have 
acquired an additional factor $Z_{\lambda\mu}$. These corner plaquettes are 
called twist-carrying and the change of variables has absorbed the twist in 
the action, by multiplying these plaquettes by the appropriate center element 
(the action involves the real part of the plaquette variables {\em after} this 
multiplication). The location of the twist-carrying plaquette is arbitrary, as 
one is free to choose the boundary of the box used for defining the torus. 
Alternatively, the twist-carrying plaquette can be moved around by a {\em 
periodic} gauge transformation. It corresponds to the non-Abelian analogue of 
a Dirac string, and is at the heart of 't Hooft's definition of magnetic flux 
for non-Abelian gauge theories~\cite{THBoper,THT4}. Thus, twist is introduced 
by the trivial modification of the weights of the plaquettes in terms of 
multiplication with appropriate center elements and causes no computational 
overhead. 

Note that we have just shown that if $Z_{\mu\nu}=1$ for all $\mu$ and $\nu$, in
a suitable gauge the links can be chosen periodic without changing the weights 
of the plaquettes. In the continuum, however,  there remains an obstruction
in making the 
gauge field periodic when the topological charge of the configuration is 
non-trivial~\cite{PCMP,TonRev}. This shows that on the lattice, only the center
charges are 
unambiguously defined. Interestingly this includes configurations~\cite{MaTo} 
that in the continuum would be assigned a non-trivial fractional Pontryagin 
index~\cite{THT4} (so-called twisted instantons).

To understand what the effect of the twist is on the holonomy, we use the 
observation that the presence of the $Z_N$ flux can be measured by taking a 
Polyakov loop in the $a^\lambda$ direction, which when translated over 
a period in the $a^\mu$ direction picks up a factor $Z_{\mu\lambda}$. 
\be
P_\lambda(x)=\frac{1}{N}{\Tr~P}\exp\left(\int_0^1A_\lambda(x+sa^\lambda)ds
\right)\Omega_\lambda(x),\quad P_\lambda(x+a^\mu)=Z_{\lambda\mu}P_\lambda(x).
\ee
There are various ways to see this~\cite{PThe,UsCo}, but becomes most evident
when `pulling' the loop over the twist-carrying plaquette.
For $SU(2)$ this means that the Polyakov loop is anti-periodic in case the
twist is non-trivial. In particular for $Z_{0i}=-1$, $P_0(\vec x)$ is
anti-periodic in the $x_i$-direction. As  we increase the size of the
spatial torus it is natural to expect  that the self-dual configuration would 
approach a caloron solution.  Then  $P_0(\vec x)$  would
approach  a constant at spatial infinity.   
 This is only compatible with the anti-periodicity implied by the 
non-trivial time-twist when $P_0(\vec x)\rightarrow 0$ for $|\vec x|\rightarrow
\infty$, forcing $\omega=\quart$ and thus non-trivial holonomy. This therefore 
provides a sure way of obtaining caloron solutions with non-trivial holonomy 
on the lattice, which at high temperature gives rise to two constituent 
monopoles, albeit in this case of equal mass. 

Since the twist in the time direction forces the constituent monopoles to have
equal mass, the lattice corrections to the value of the action (which depend
on the shape of the
configuration~\cite{UsCo}) are affected only by the separation of the two
constituents (in the next section we will encounter the situation where the mass
ratio is affected by the cooling). This allows to manipulate the positions
of the two lumps by using the tool of cooling with modified actions.
This can be implemented~\cite{UsCo} by using a lattice action that combines
the traces of the $1\!\times\!1$ and $2\!\times\!2$ plaquettes. The two 
couplings are fixed in terms of the parameters multiplying the leading 
(continuum) and next to leading ($a^2$) terms in the expansion of the lattice
action  in powers of the lattice spacing $a$. The $a^2$ term
is given by a unique dimension six operator, and its coefficient
is called $\eps$ (it is trivial to incorporate the twist-carrying plaquettes
also in these modified actions). Wilson's action corresponds to
 $\eps\!=\!1$. The choice $\eps < 1$ is known as
over-improvement, whereas improved cooling is performed by choosing $\eps\!=\!0$.
In this last case the lattice and continuum action differ only by corrections
of order $a^4$. For that reason, we will choose $\eps\!=\!0$ whenever we compare
with the analytic infinite volume continuum caloron
solution. However, unlike for the continuum action, the value of the
$a^2$ operator 
depends on the position of the constituent monopoles, 
and therefore we can  use other values of  $\eps$  to alter these positions. 
Cooling with the 
Wilson action has the effect of driving the constituent monopoles together, 
since the Wilson action is decreased   with respect  to the continuum when the 
field strength has a  larger gradient~\cite{UsCo}. Once the two lumps merge, 
and can no longer be distinguished from an instanton (at which point the 
solution will no longer be static), it follows the usual fate of an instanton 
under prolonged cooling with the Wilson action: At some point it falls through 
the lattice~\cite{Teper}. (For cooling histories see fig.~3).

Over-improved cooling has the effect of pushing the two constituent monopoles 
apart.
One can  speed-up the rate at which monopoles separate 
by decreasing  $\eps$. Apriori it is not clear if, when the lumps are 
maximally apart, the solution will not be affected significantly by the 
boundary conditions. This will partly depend on the ratio $L/\beta$, but we 
find for $L=4\beta$ that these effects are rather small. 

\begin{figure}[htb]
\vspace{4.8cm}
\includegraphics{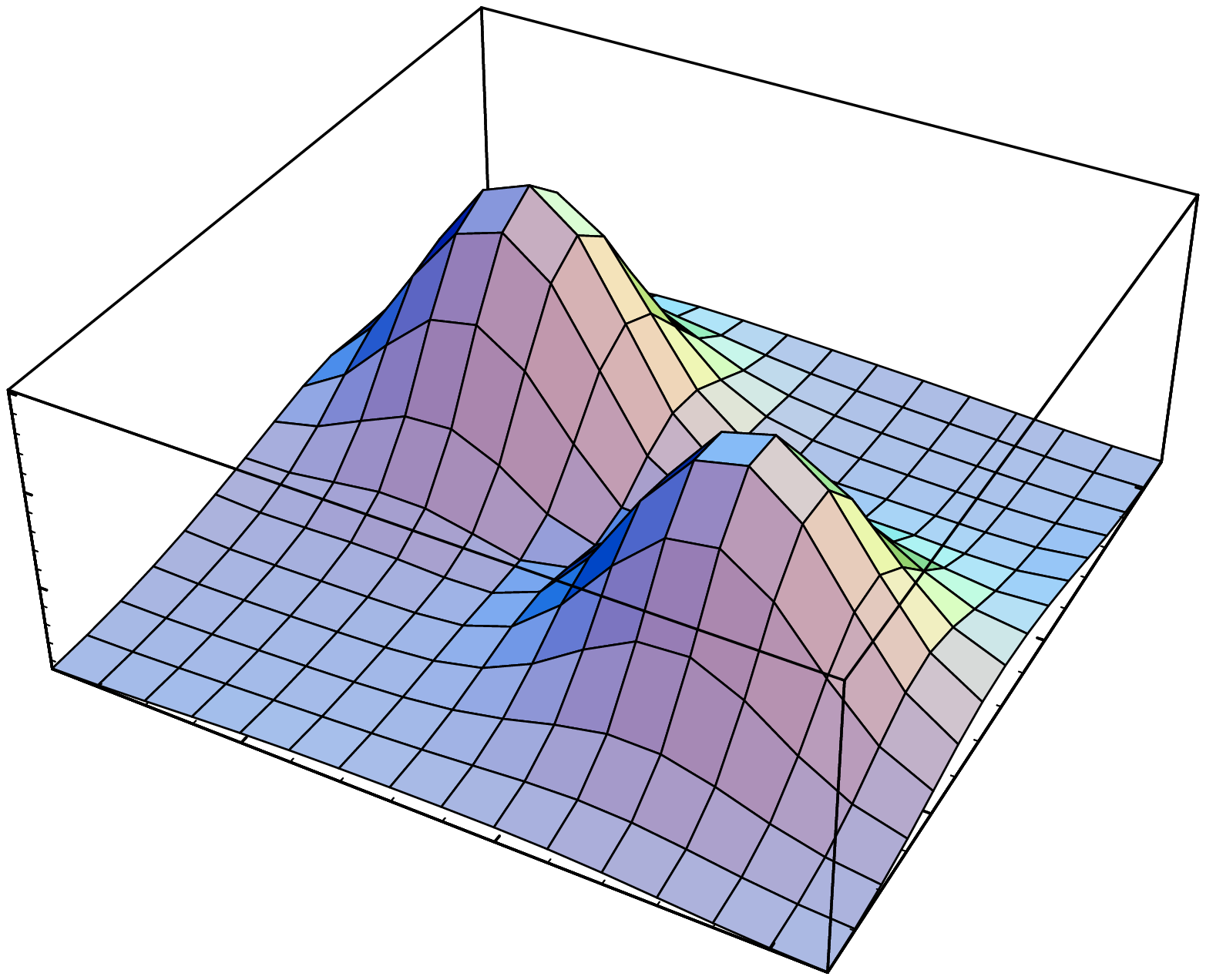}
\includegraphics{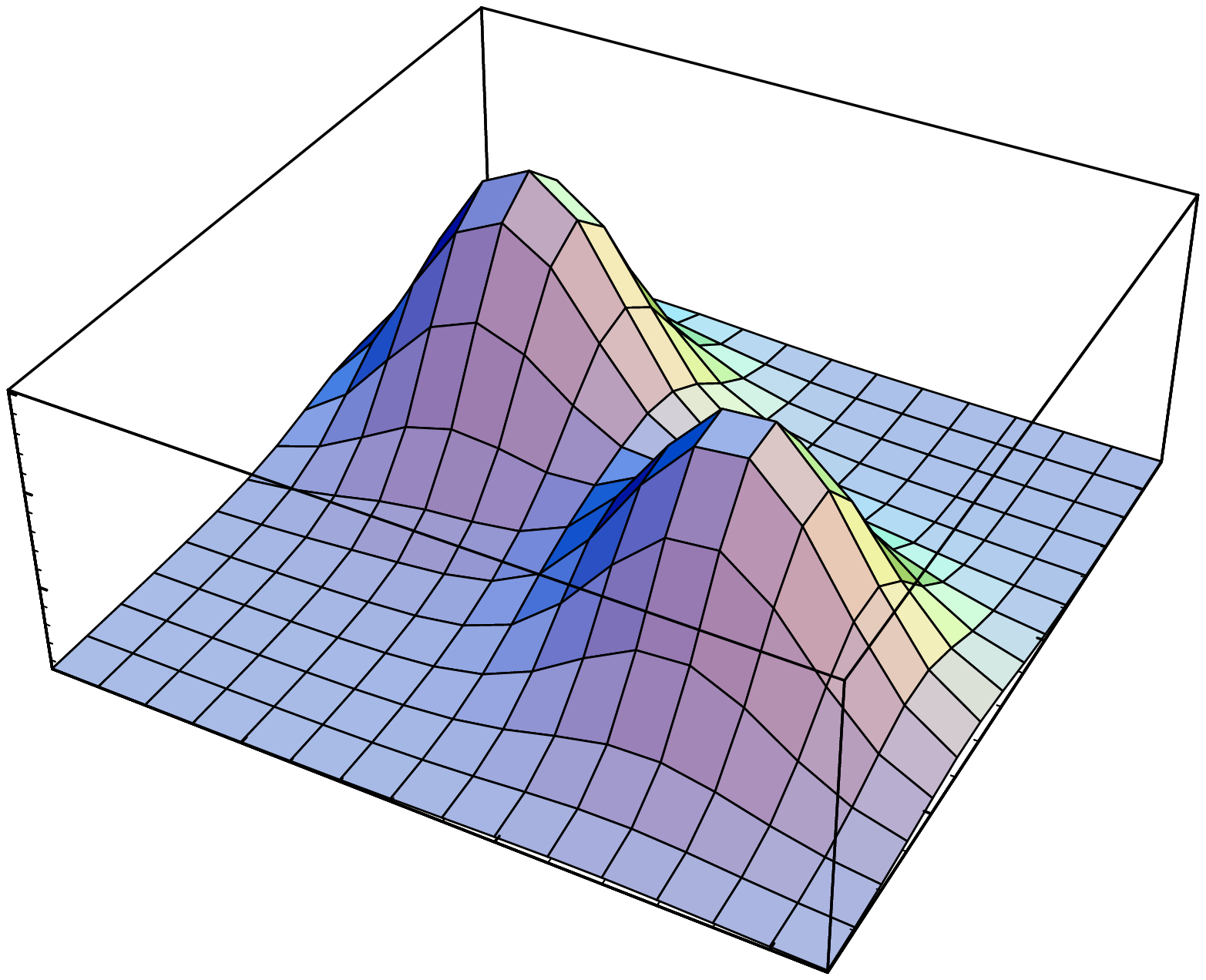}
\caption{Lattice caloron profiles (left) on a $16^3\times 4$ lattice for 
$\vec k=(1,1,1)$, created with improved cooling ($\eps=0$). The total 
action is $1.000185\times8\pi^2$. Vertically is plotted $\log(1+s/3)$, with 
$s$ the action density at the lattice site (after clover averaging). The profile 
fits well to the analytic caloron solution (shown on the right at $y\!=\!t\!=\!0$) 
with $\omega=\quart$ and constituents at $\vec y_1=(2.50,0.12,0.95)$ and 
$\vec y_2=(1.38,-0.24,2.67)$, in units where $\beta=1$ (or $a=\quart$) 
and the left most lattice point corresponding to $x=z=0$.}
\end{figure}

In figure 1 we give an example of a caloron configuration with well separated
constituents on a $16^3\!\times\!4$ lattice with $\vec k=(1,1,1)$, initially 
generated by cooling with the Wilson action, switching to improved cooling 
to reduce lattice artifacts. Shown is the action density $s$. We see that 
the agreement with the infinite volume analytic result is very good, with 
the action peaks for the lattice result somewhat lower (this feature is 
somewhat suppressed by plotting $\log(1+s/3)$, rather than $s$). The total 
action of this static lattice configuration is very close to the required 
continuum value $8\pi^2$. An example of a non-static configuration with
overlapping constituents will be presented below (see fig.~5). There 
seems no doubt that a continuum solution with this constituent monopole 
structure should exist on the time-twisted torus. 

\section{The case of space-twist}

When both space  and time twists are non-trivial and $\vec k\cdot\vec m\!\neq\!
0~\mod~N$ (called non-orthogonal twist), the minimum of the action corresponds 
to a so-called twisted instanton with fractional charge. Unlike the integer 
charge instantons, these twisted instantons can not fall through the lattice.
Their scale is fixed, only their position is a free parameter. This 
was used in the past~\cite{MaTo} to find accurate lattice results using 
ordinary cooling ($\eps\!=\!1$). At high temperatures such a twisted instanton 
becomes static and represents a single BPS monopole on $T^3$. The twist 
allows for non-zero charge in the box. As discussed in the previous section it 
also gives rise to a holonomy characterised by $\omega=\quart$. Indeed, we were 
able to fit the finite temperature twisted instanton (in a sufficiently large 
volume) to one of the constituent monopoles of the caloron at $\omega=\quart$ 
(when placing the other constituent at a sufficiently large separation). In 
the appropriate limits both become ordinary BPS monopoles with mass 
$4\pi^2/\beta$.

\begin{figure}[htb]
\vspace{6.7cm}
\includegraphics{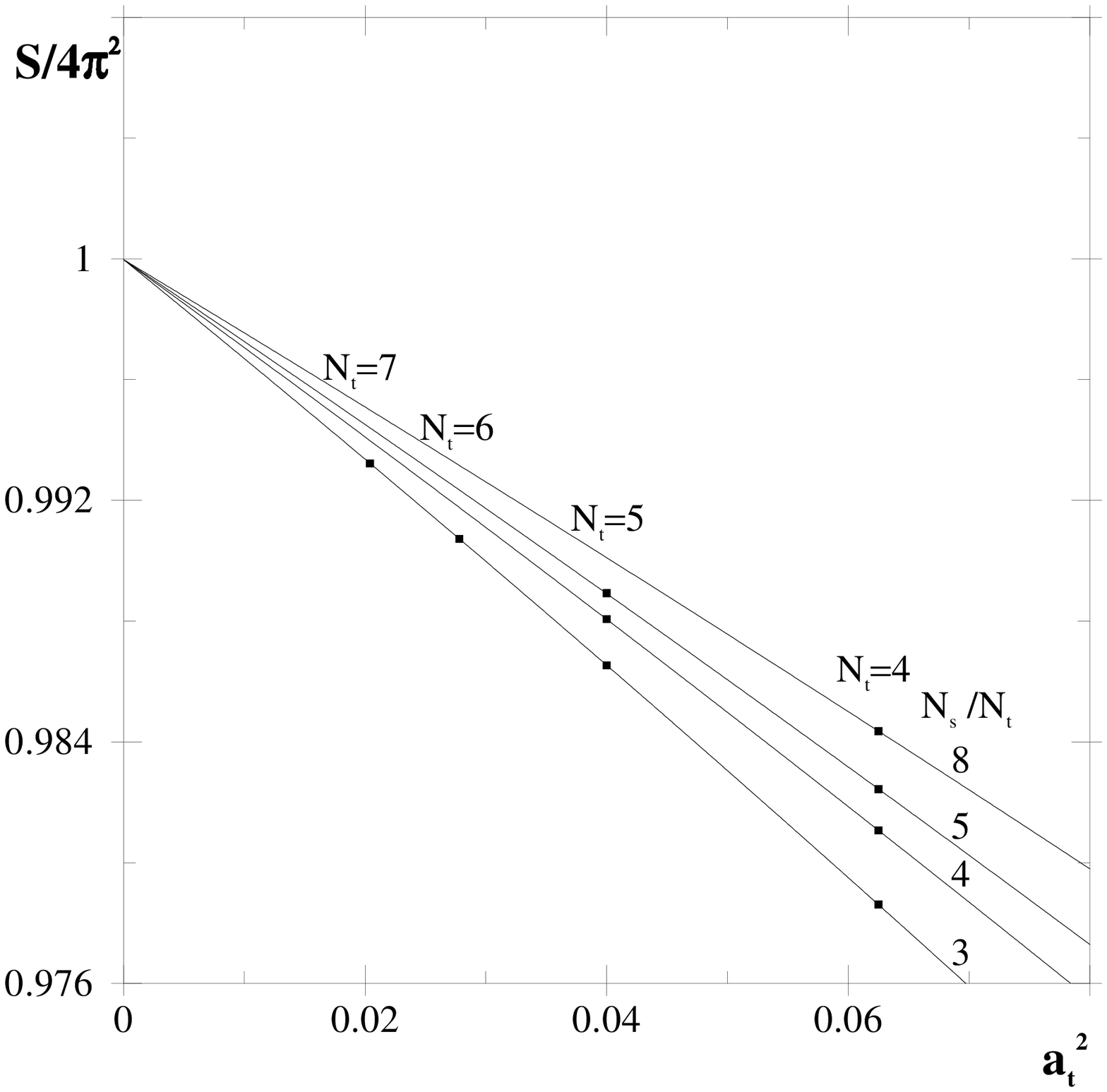}
\includegraphics{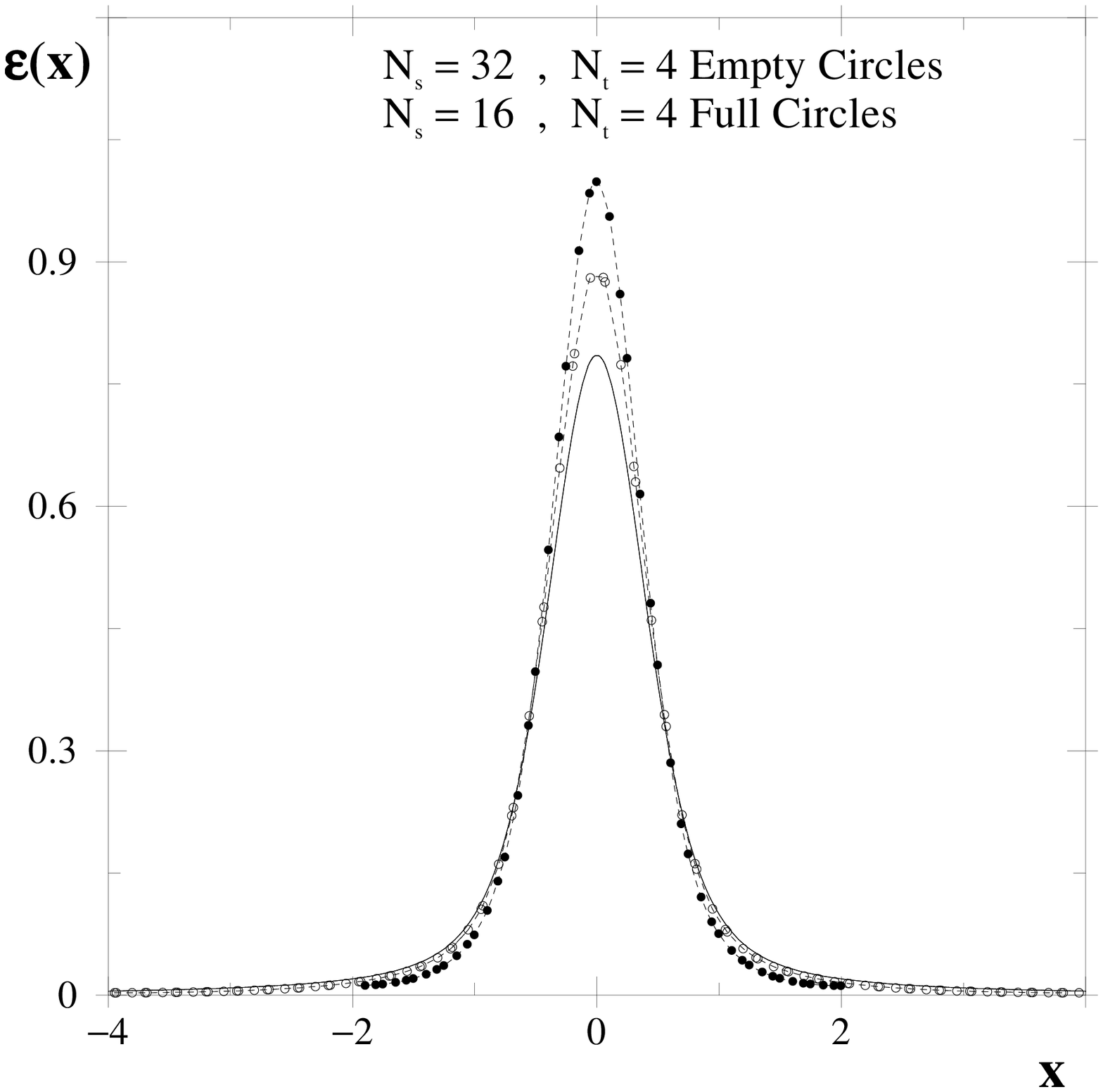}
\caption{Lattice minimum action for various sizes (left), including the best
fit to eq.~(5) with: $S_0=0.999975(4)$, $b=0.19950(7)$, $c=0.3107(3)$ and 
$d=0.0844(5)$. Also shown is the lattice action profile ${\cal E}(x)$, obtained
by summing the action density over all but one of the spatial coordinates, 
as compared to the infinite volume analytic BPS result (right).}
\end{figure}

Now we will show the type and size of finite volume and lattice artifact effects. In fig.~2 (left)
we display a plot of the minimum lattice (Wilson) action for lattices with different
space and time extensions, $N_s^3\!\times \!N_t$ and twist
$\vec{k}\!=\!\vec{m}\!=\!(1,1,1)$. In the given range, $N_t=4$---$7$ and 
$N_s=16$---$32$, deviations from the continuum result are of the order of 
a few percent. However, the pattern of deviations from the continuum value 
is well understood. If we set $a_t=1/N_t$ and $a_s=1/N_s$, the value of the 
lattice action can be fitted with great accuracy to a formula:
\be
S/4\pi^2=S_0-b a_t^2-b a_s^2-c a_t a_s-d(a_t+a_s)^4 .
\ee
The extrapolated value of the continuum action matches $4\pi^2$ to a precision
of a few parts in $10^5$. Notice also that the extrapolation shows the
existence of a self-dual continuum solution for any value of the ratio
$L/\beta\!=\!N_s/N_t$. Furthermore, the lattice correction to the action 
decreases in absolute value with the ratio $N_s/N_t$, consistent with the 
statement made before that Wilson's action decreases with decreasing 
separation of the monopoles, since in this case $N_s$ plays the role of 
the separation between lumps (the periodic mirrors).

To measure finite volume corrections, we performed improved cooling ($\eps\!=\!0$,
to minimise lattice corrections) for $N_s\!=\!16$ and $32$.
In this case, the values of the minimum lattice
action attained are of the order  $S/4\pi^2\!=\!1.0001(1)$. In fig.~2 (right) we
compare the $x$-profiles ${\cal E}(x)$ obtained from the lattice minimum
action configuration with the corresponding one for the BPS monopole.
The $x$-profile
is the integral of the action density over all but the $x$ coordinate. This
quantity has smaller errors and is less sensitive to the lattice discretisation
than the action density itself. From the figure we see how the lattice
profiles approach the infinite volume BPS monopole profile. The slow
convergence is due to the powerlike Abelian tail of the BPS monopole 
(in contrast to the exponential tail found for other cases~\cite{R2xT2}).

Interestingly, an exact caloron solution with equal-size constituents ($\omega=
\quart$) on the twisted torus can be constructed by gluing two twisted
instantons together, starting from the $Q=\half$ solution defined by $\vec k=
\vec m=(1,0,0)$. Gluing two boxes in the $y$- or $z$-directions preserves 
$\vec k$, but reduces $\vec m$ to the trivial value (since $n_{\mu\nu}$ is 
defined modulo 2 for $SU(2)$). This exact solution corresponds to the situation 
studied in the previous section. Instead, gluing two boxes in the 
$x$-direction removes the time-twist, but preserves the space-twist. The same 
twist results when gluing the two boxes in the time-direction. In the first
case we have an exact solution on a space-twisted torus with equal size 
constituents (corresponding to $\omega=\quart$) at maximal separation
in the direction of the twist, whereas in the second case the static nature of 
the finite temperature solution simply leads to doubling the mass of the 
monopole. Therefore this solution corresponds to an exact caloron solution on 
a space-twisted torus with trivial holonomy (the other constituent monopole 
is massless).

\begin{figure}[htb]
\vspace{5.4cm}
\includegraphics{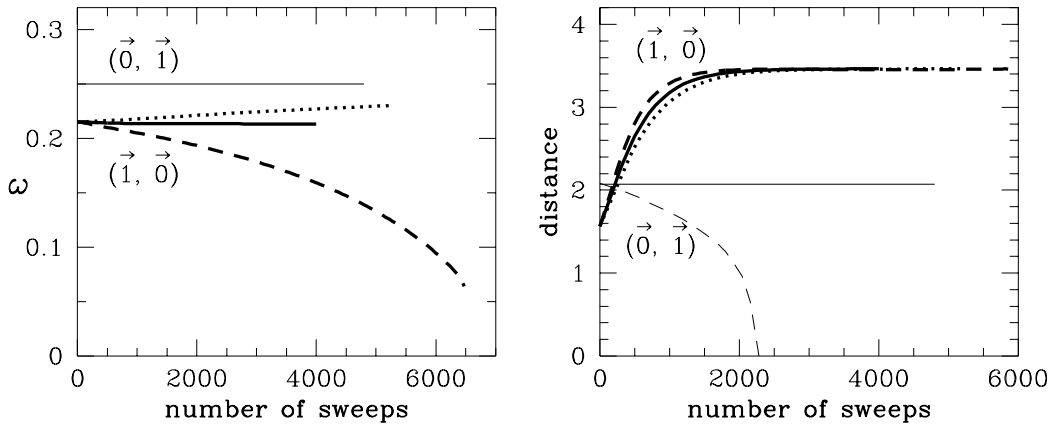}
\caption{Cooling histories for $(\vec m,\vec k)=(\vec 0,\vec 1)$ and
$(\vec 1,\vec 0)$ (resp. thin and fat curves), where $\vec 1\equiv(1,1,1)$, on
lattices of size $16^3\times4$. Solid, dashed and dotted curves are for resp.
$\eps=0,1,-1$ cooling. For $(\vec 0,\vec 1)$, where $\omega\equiv\quart$, the
two $\omega$--curves cannot be distinguished.}
\end{figure}

We have also performed lattice studies on a space-twisted torus, with 
$\vec k=\vec 0$, which allowed us to probe the constituent monopole 
mass ratios, by a subtle use of the cooling procedure. It can be proven that 
without twist there are {\em no regular} charge one instanton solutions on a 
torus~\cite{BrvB}, but for any non-trivial twist an 8 dimensional space of 
regular solutions exists~\cite{Taub,Braam}. Part of this parameter space comes 
about by gluing a localised instanton to the unique curvature free background 
supported by the twist. The eight parameters are given by scale, space-time 
position and so-called attachment parameters, that describe the gauge 
orientation of the localised instanton relative to the fixed curvature free 
background. For $\vec m\neq\vec 0$ we find that the magnetically charged 
constituent monopoles, superposed on this non-Abelian magnetic flux background,
experience an additional force that repels them as far as the finite volume 
allows. The presence of this force is evident from the fact that 
under prolonged cooling in all cases, $\eps=1,0,-1$, the separation between 
the two constituent monopoles was increasing and that their centers lined up 
with the direction of $\vec m$. Once the constituent monopoles are placed at 
their maximal separation, further cooling with the Wilson action ($\eps\!=\!1$) 
leads to action shifting from one to the other peak, driving the constituent 
monopole mass ratios away from equal masses. Once one of the masses has 
decreased to zero, the scale parameter of the remaining (deformed) instanton 
configuration can shrink, resulting in the usual fate of falling through the 
lattice under prolonged cooling with the Wilson action. For over-improvement 
the effect is opposite, and the masses are pushed to equal values. The `force' 
---due to lattice artifacts--- changing the value of $\omega$ can be neglected 
for $\eps\!=0$ cooling. We summarise the behaviour under cooling in 
fig. 3, by showing the distance between the peak locations and $\omega$ 
(estimated by equating $(\half\!-\!\omega)^4 /\omega^4$ to the ratio of the peak 
heights) as a function of the number of cooling sweeps. Shown are the histories
for $\vec m\!=\!(1,1,1)$ at $\eps\!=\!-1,0,1$ and for $\vec k\!=\!(1,1,1)$ at 
$\eps\!=\!0,1$.

\begin{figure}[htb]
\vspace{9.7cm}
\includegraphics{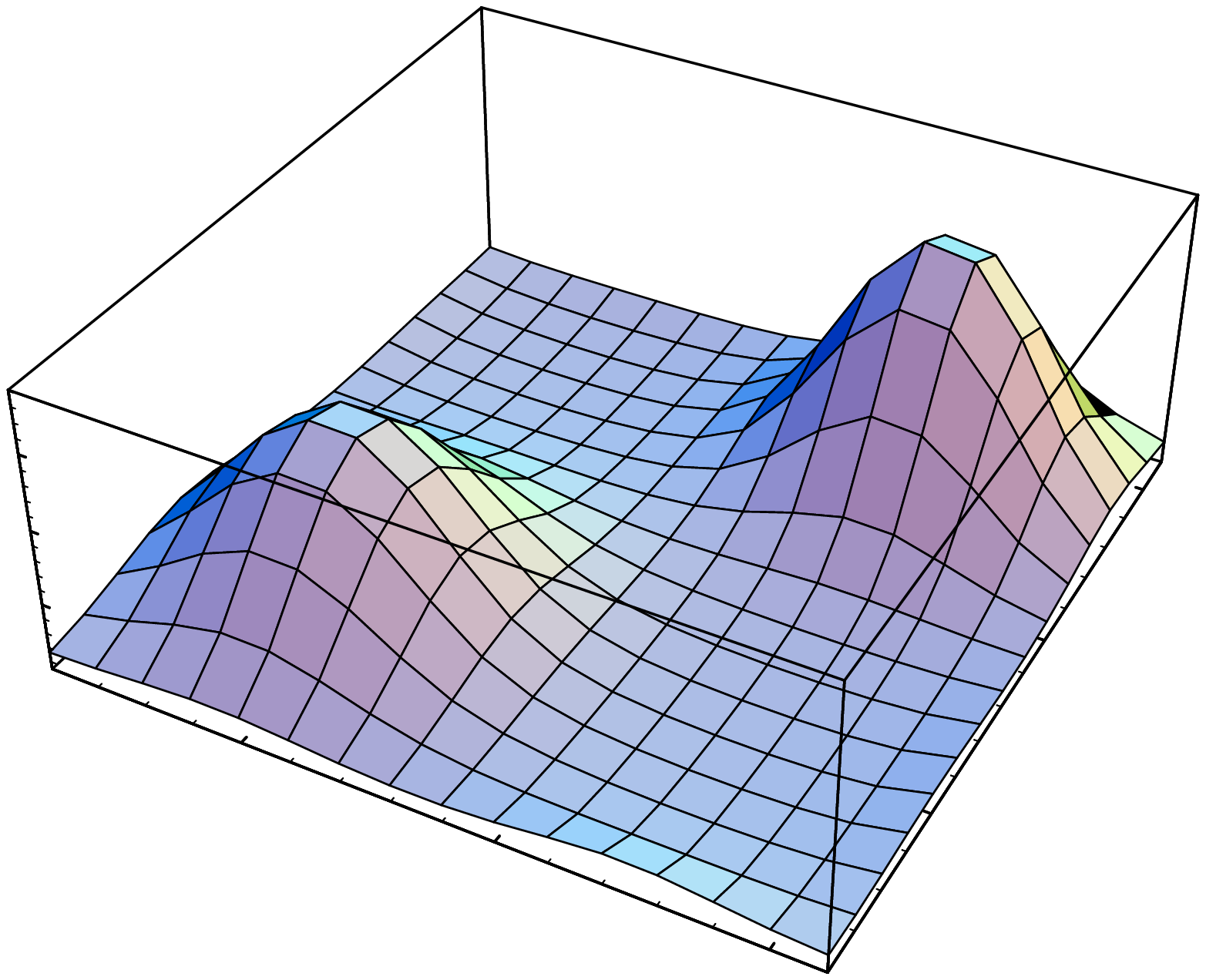}
\includegraphics{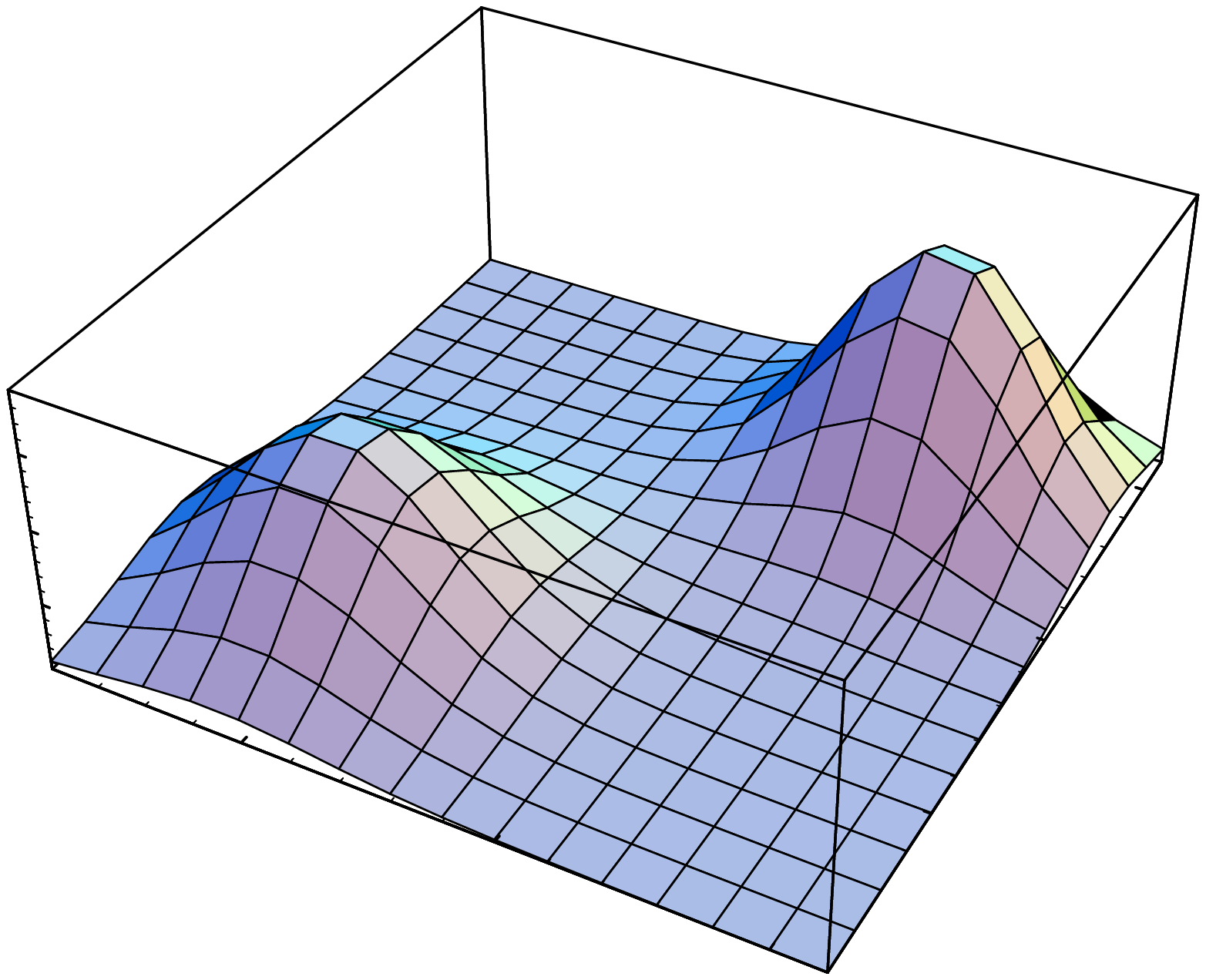}
\includegraphics{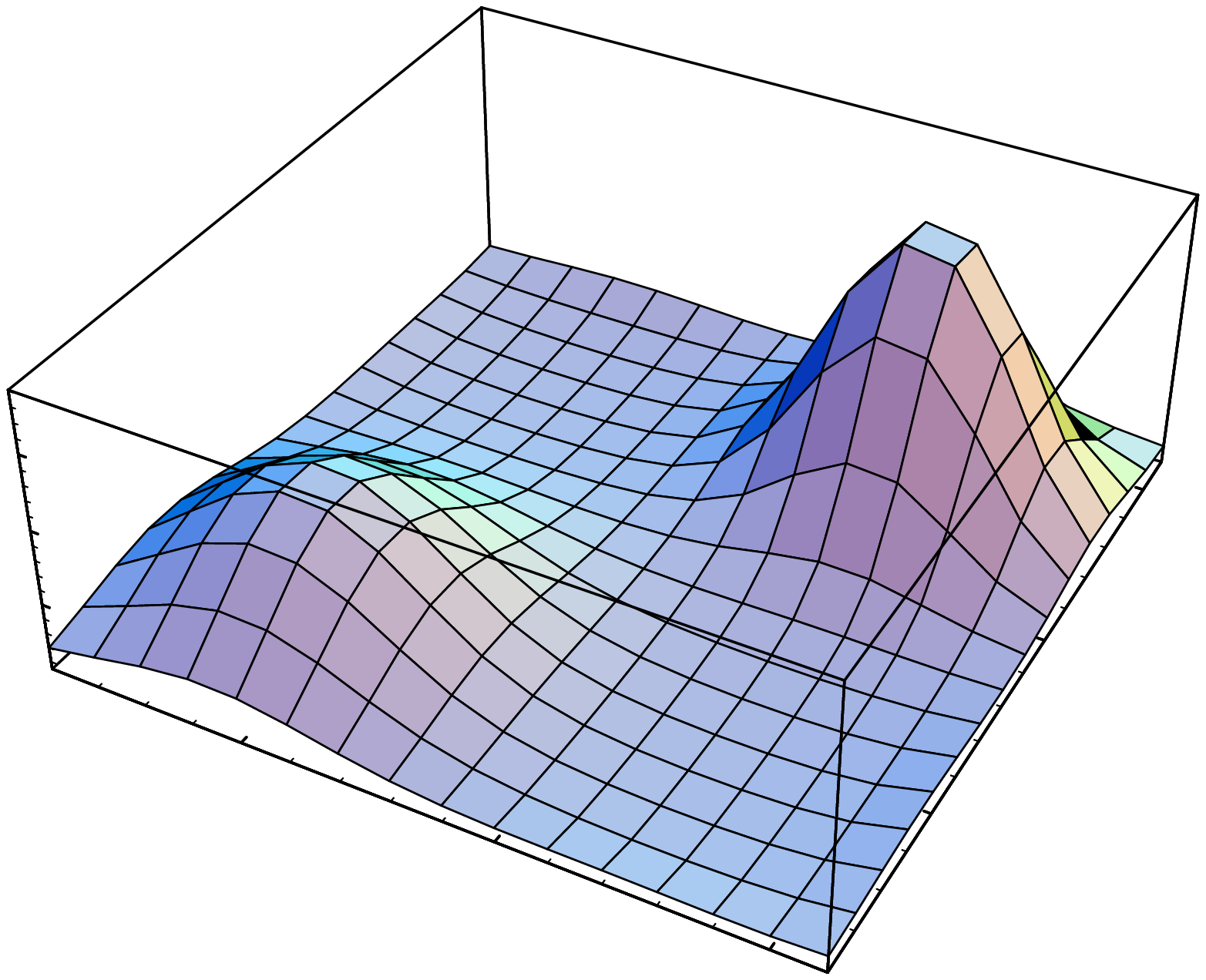}
\includegraphics{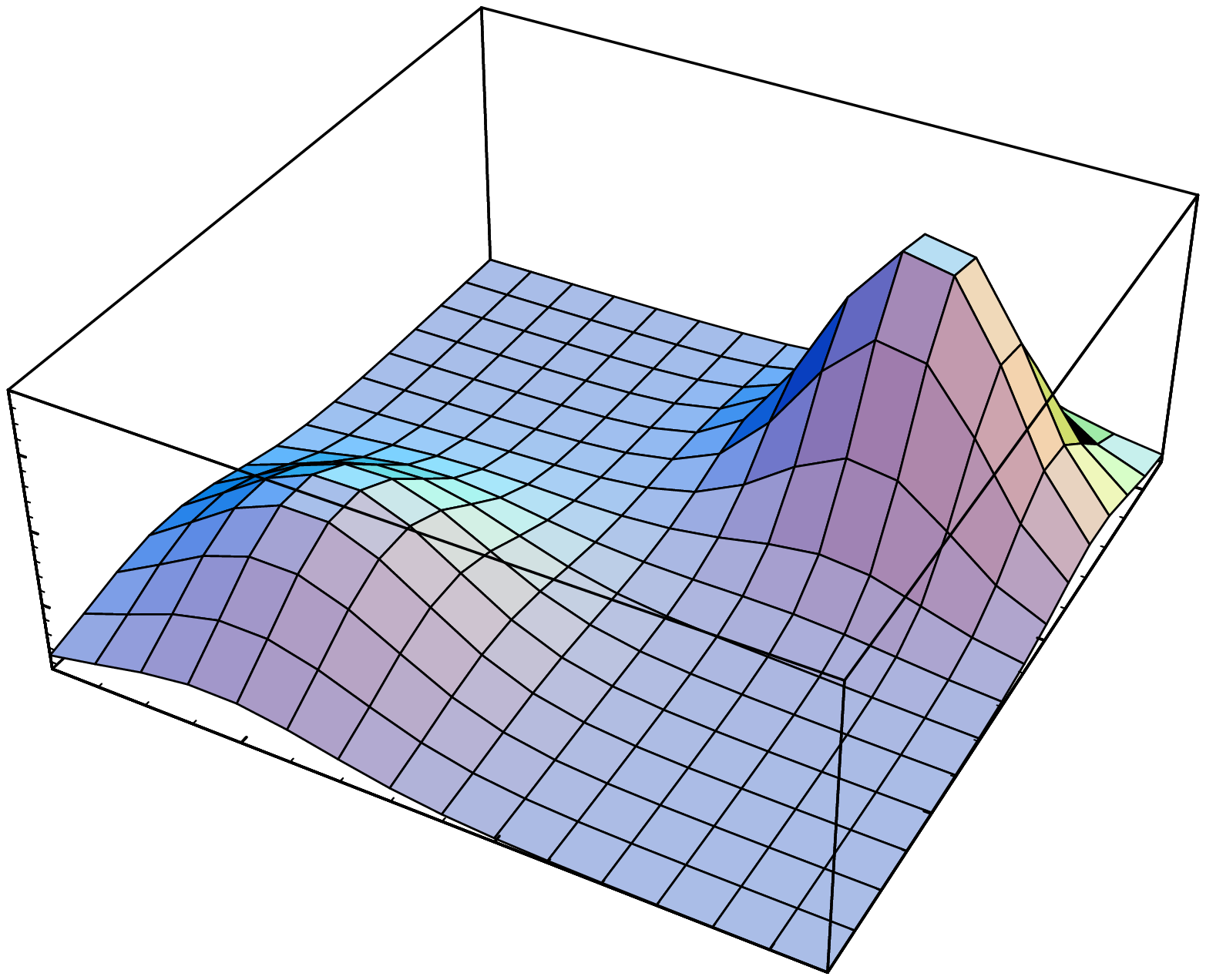}
\caption{Lattice caloron profiles (left) for two configurations
on a $16^3\times 4$ lattice for $\vec m
=(1,1,0)$, created with improved cooling ($\eps=0$) - after manipulating with
$\eps=\pm1$ cooling to obtain the desired mass ratios. The total actions are
$1.000155\times 8\pi^2$ (top) and $1.000001\times 8\pi^2$ (bottom). Vertically
is plotted $\log(1+s/3)$, with $s$ the action density at the lattice site
(after clover averaging). The profiles fit well to the analytic caloron
solutions (shown on the right at $y=t=0$) with top: $\omega=0.210$ and
constituents at $\vec y_1=(1.04,-0.08,0.86)$ and $\vec y_2=(3.05,-0.09,2.85)$,
and bottom: $\omega=0.175$ with constituents at $\vec y_1=(0.85,-0.06,0.85)$
and $\vec y_2=(2.85,-0.06,2.85)$, all in units where $\beta=1$ (or $a=\quart$) and
the left most lattice point corresponding to $x=z=0$.}
\end{figure}

That we can have solutions that are characterised by arbitrary mass ratios of the
constituent monopoles is also illustrated in figure 4, which represents two
values for the parameter $\omega$, comparing the finite volume configurations
obtained from improved cooling to the analytic infinite volume 
caloron solutions with non-trivial holonomy. We see again that the agreement 
is very good (and will improve for increasing $L/\beta$), with the peaks for 
the lattice result now somewhat higher as compared to the infinite volume 
results. 

\begin{figure}[htb]
\vspace{9.7cm}
\includegraphics{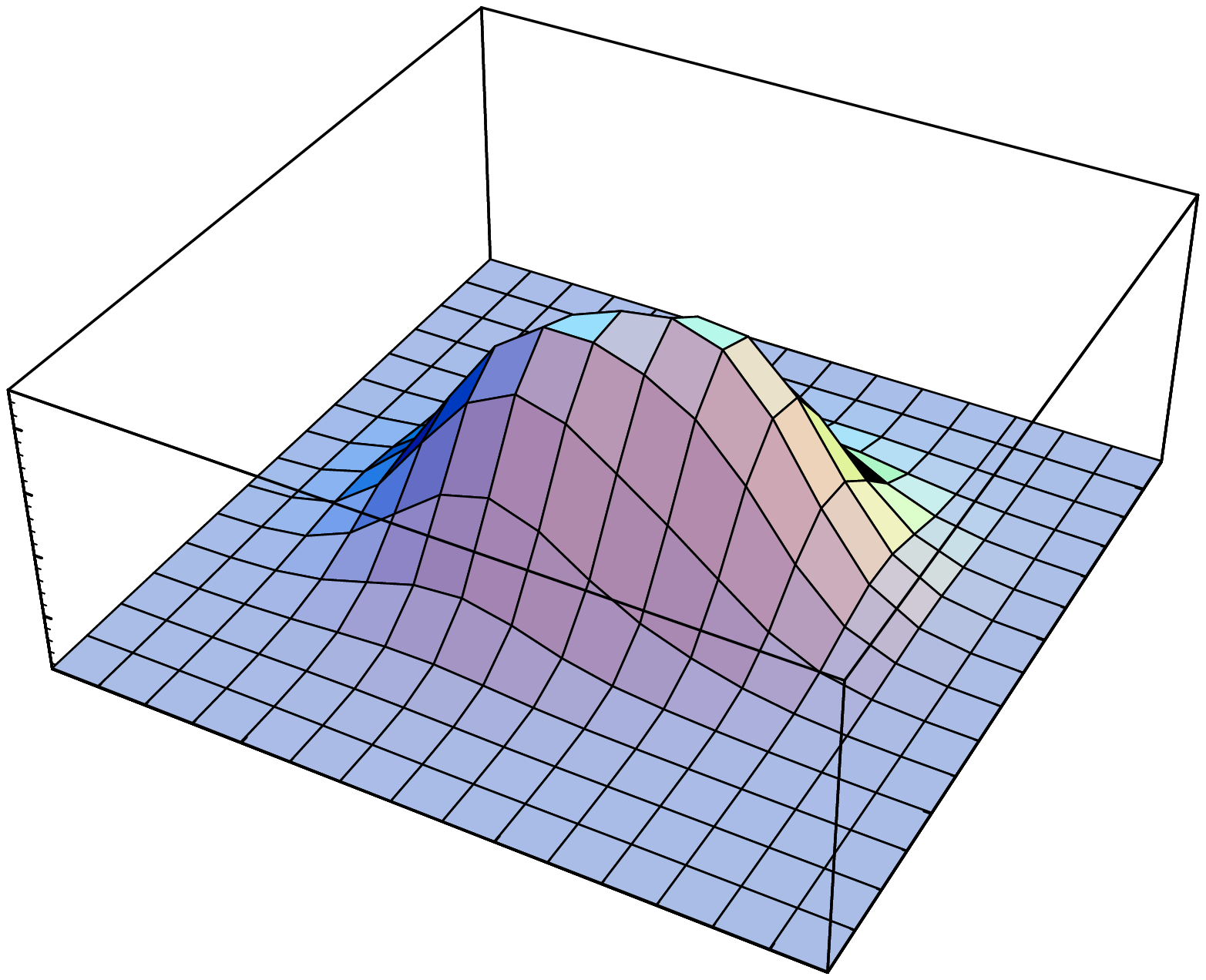}
\includegraphics{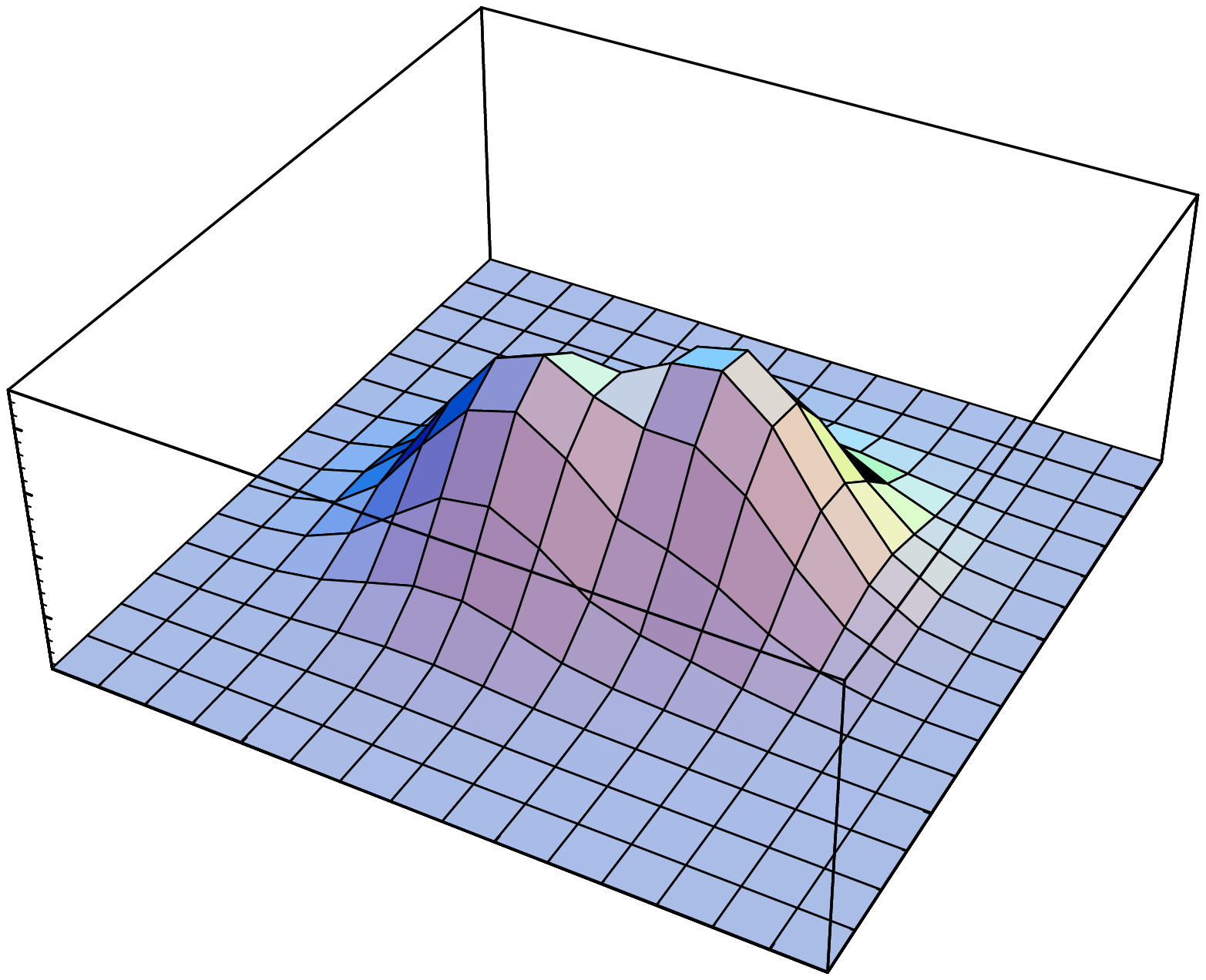}
\includegraphics{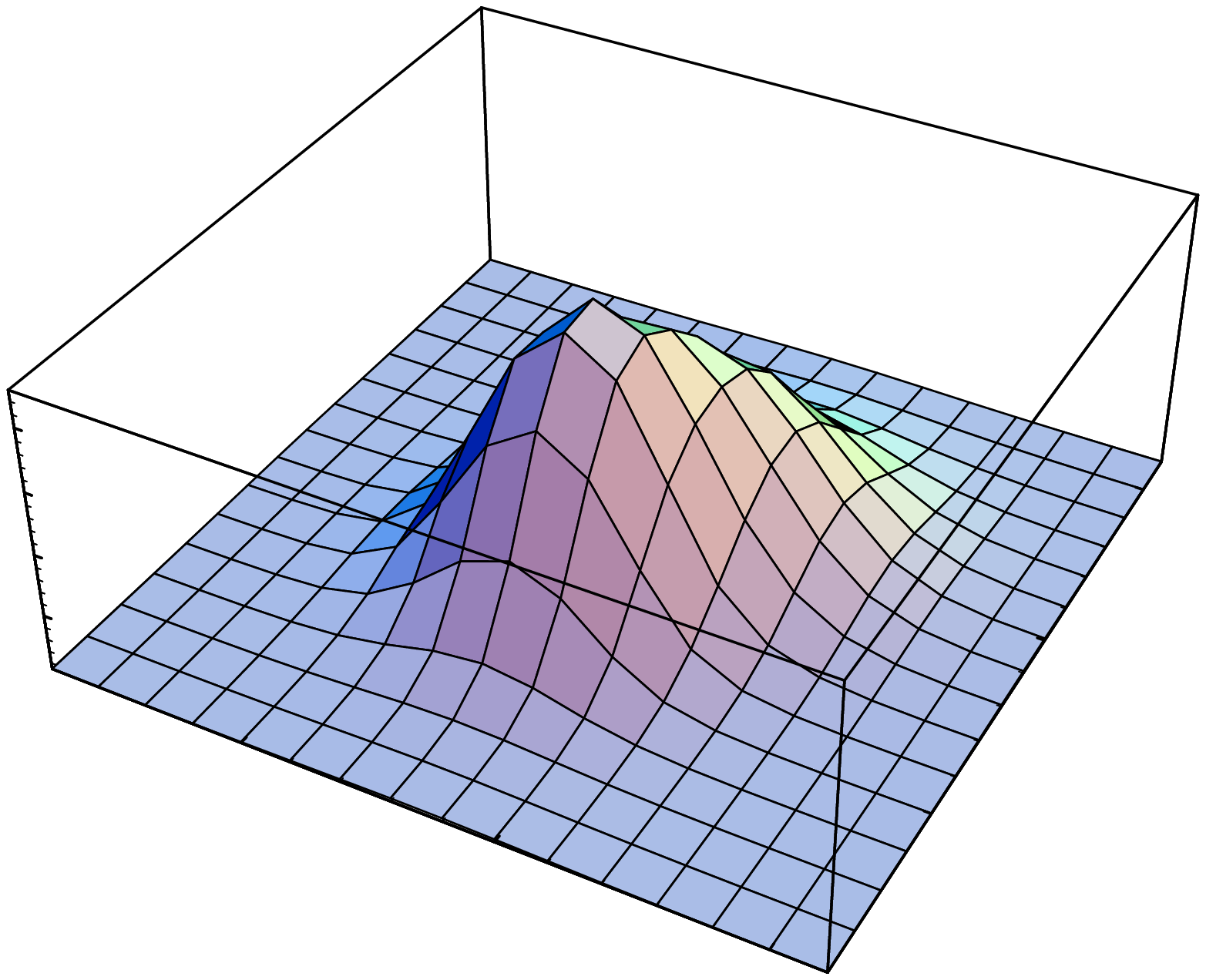}
\includegraphics{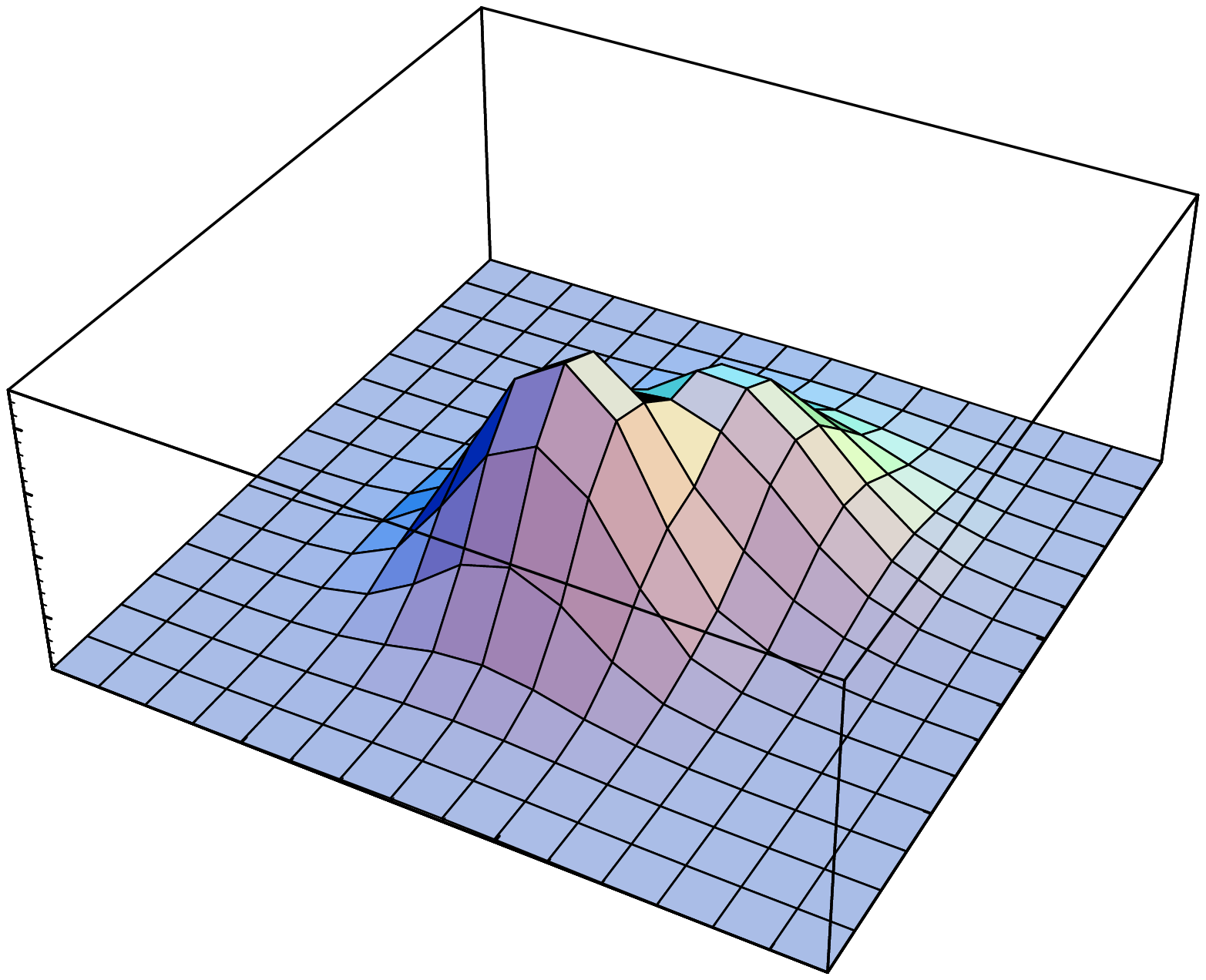}
\caption{Caloron profile for two configurations (left and right)  on a $16^3\times 4$ lattice 
for $\vec m=\vec 0$ 
obtained with improved ($\eps=0$) cooling. Vertically is plotted
$\log(1+s/3)$, with $s$ the action density at the lattice site (after clover
averaging). The top plot  corresponds to the plane   $y=t=0$
 and the bottom one to   $y=0$, $t=\half$.
Left: $\vec k=(1,1,1)$ with total action $1.000016\times 8\pi^2$, 
$\omega=\quart$, $t_0=-0.125$, $\vec y_1=(1.75,-0.15,1.43)$ and $\vec y_2=
(2.20,0.00,2.20)$. Right: $\vec k=(0,0,0)$ with total action $1.010951\times 
8\pi^2$, $\omega=0.185$, $t_0=-0.09$, $\vec y_1=(2.29,0.07,2.15)$ and $\vec y_2
=(1.57,0.08,1.73)$. All coordinates obtained from fitting to the infinite 
volume analytic solutions (not shown) are in units where $\beta=1$ 
(or $a=\quart$) 
and the left most lattice point corresponding to $x=z=0$.} 
\end{figure}

Next, we  discuss the comparison with configurations that are not static. Here
the constituents are close together and therefore there is considerable overlap.
This is illustrated in figure 5, both in  the case of  twist in time as in the
case of no twist. As mentioned previously, the presence of twist in time
($\vec{k} \neq 0$) forces $\omega=\quart$, while in the absence of twist
$\omega$ can be arbitrary. Obtaining configurations with no twist requires some
care. By cooling random configurations one ends up quickly in the trivial
vacuum configuration. Hence, it is useful to start with a $Q=1$ configuration
having  $\vec m\neq\vec 0$ obtained by cooling. Then twist is eliminated from
this configuration by setting the weights of the twist-carrying plaquettes to
their standard (untwisted) value. Additional improved cooling steps were
applied to the configuration, leading to a new  solution  still having a
non-trivial $\omega$ value.  We recall that there are no exactly self-dual
$Q=1$ solutions on the torus without twist~\cite{BrvB}. However, for solutions
well-localised inside the torus the configuration is very approximately
self-dual.
Notice, nonetheless, that this reflects itself in higher values of the minimum
lattice action. For these configurations with periodic boundary conditions
performing further cooling steps with positive  or zero $\eps$  will bring the 
constituents together and leads to the standard fate of instantons on the 
lattice. This can be stabilised by $\eps<0$, and the better the solution is 
contained within the box, the closer one can take $\eps=0$ to have a stable 
lattice solution~\cite{UsCo}. 

The differences with the analytic infinite volume caloron solutions only
show themselves by small differences in peak heights (at $t=0$) and would
not be clearly visible on the scale of figure 5. Instead, in figure 6,
we display the analytic action density profile in the $z-t$ plane,
where $z$ is the axis connecting the two constituent monopole centers.
The values of $\omega$ and the distance
of the constituent monopoles is as in figure 5. It is clear that a two-lump
structure is still visible. As a function of $z$ the constituent monopoles
are best seen for $t$ values where the density is minimal ($t=\half$). 
The logarithmic scale enhances the regions of low action densities 
in favour of those with large densities, and brings out more clearly the 
constituents.

\begin{figure}[htb]
\vspace{4.8cm}
\includegraphics{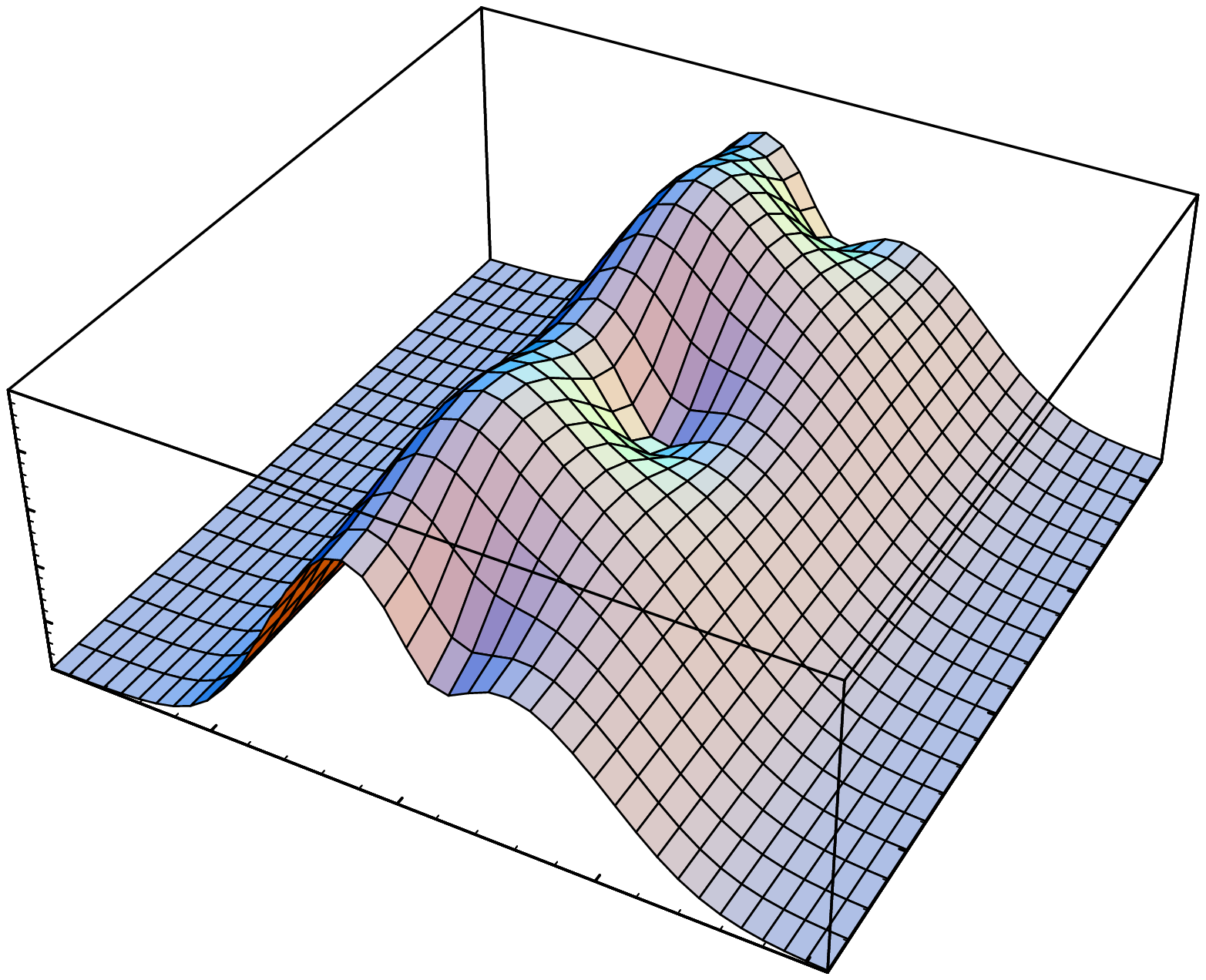}
\includegraphics{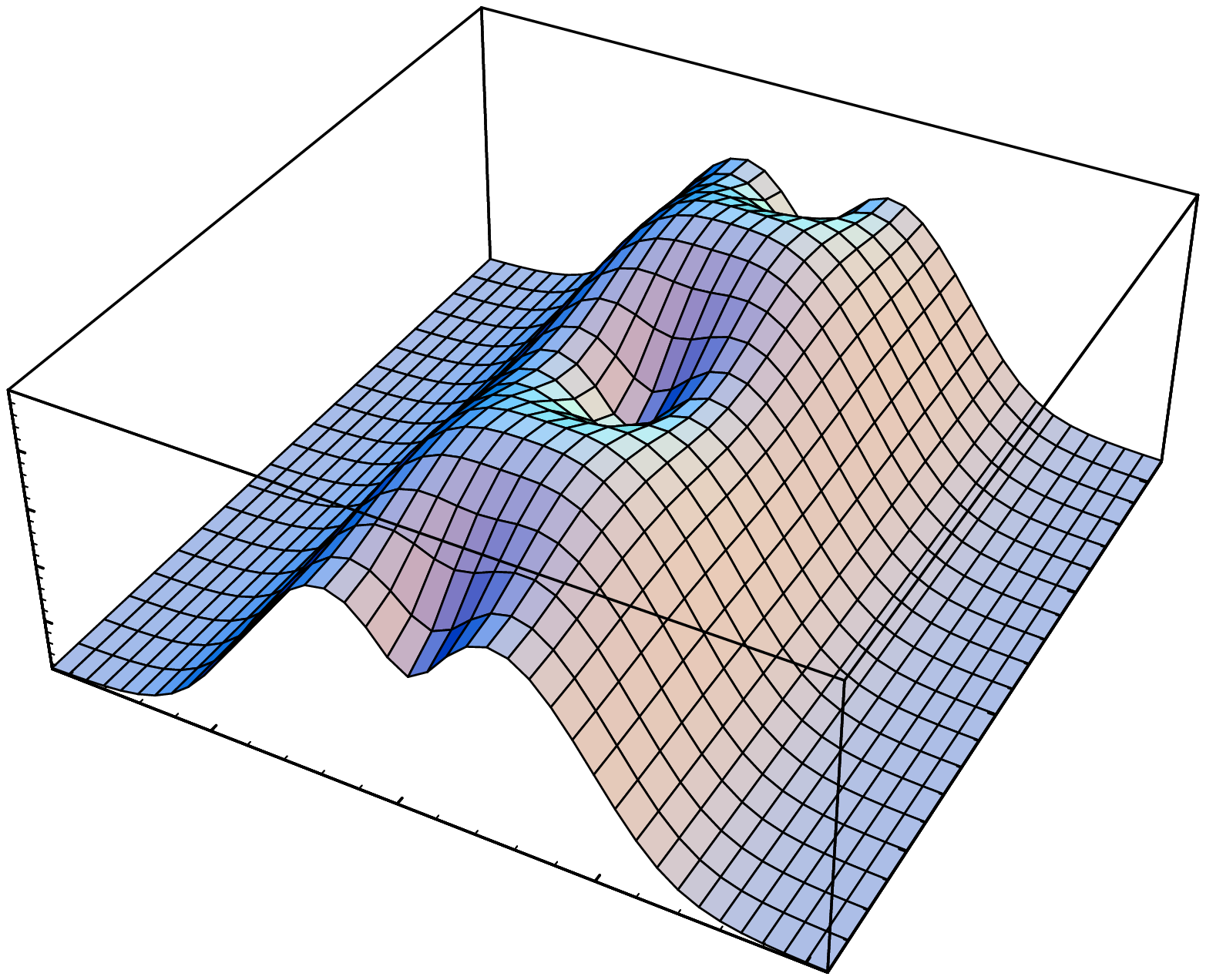}
\caption{Space-time profile for the calorons of figure 5, using the infinite
volume analytic result. Vertically is plotted $\log(1+s/3)$, with $s$ the
action density. Horizontally is plotted time ranging over two periods $\beta=1$
and space along the line connecting the two constituent monopole positions.
Left corresponds to $\omega=\quart$ and right to $\omega=0.185$.}
\vspace{5.3cm}
\includegraphics{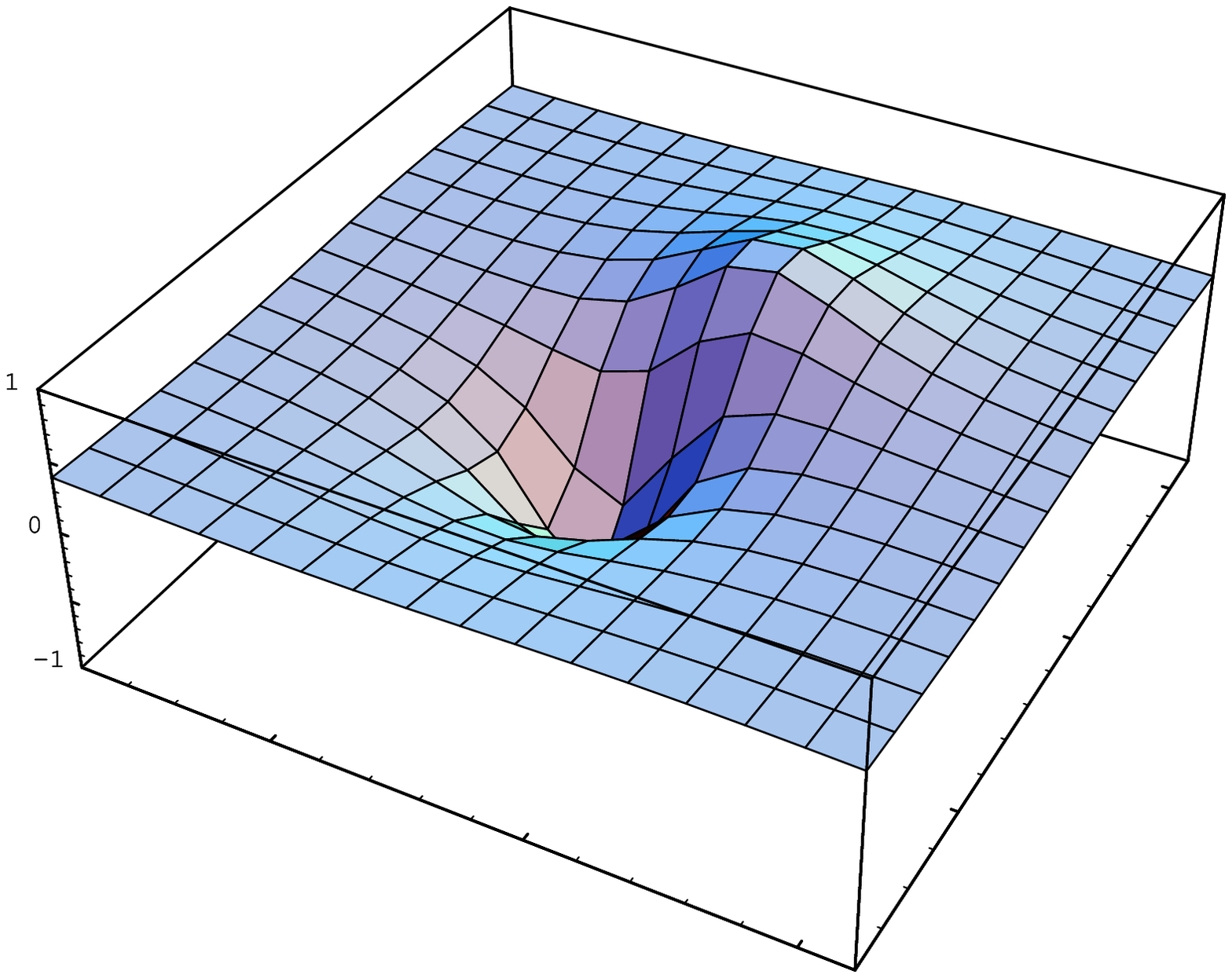}
\includegraphics{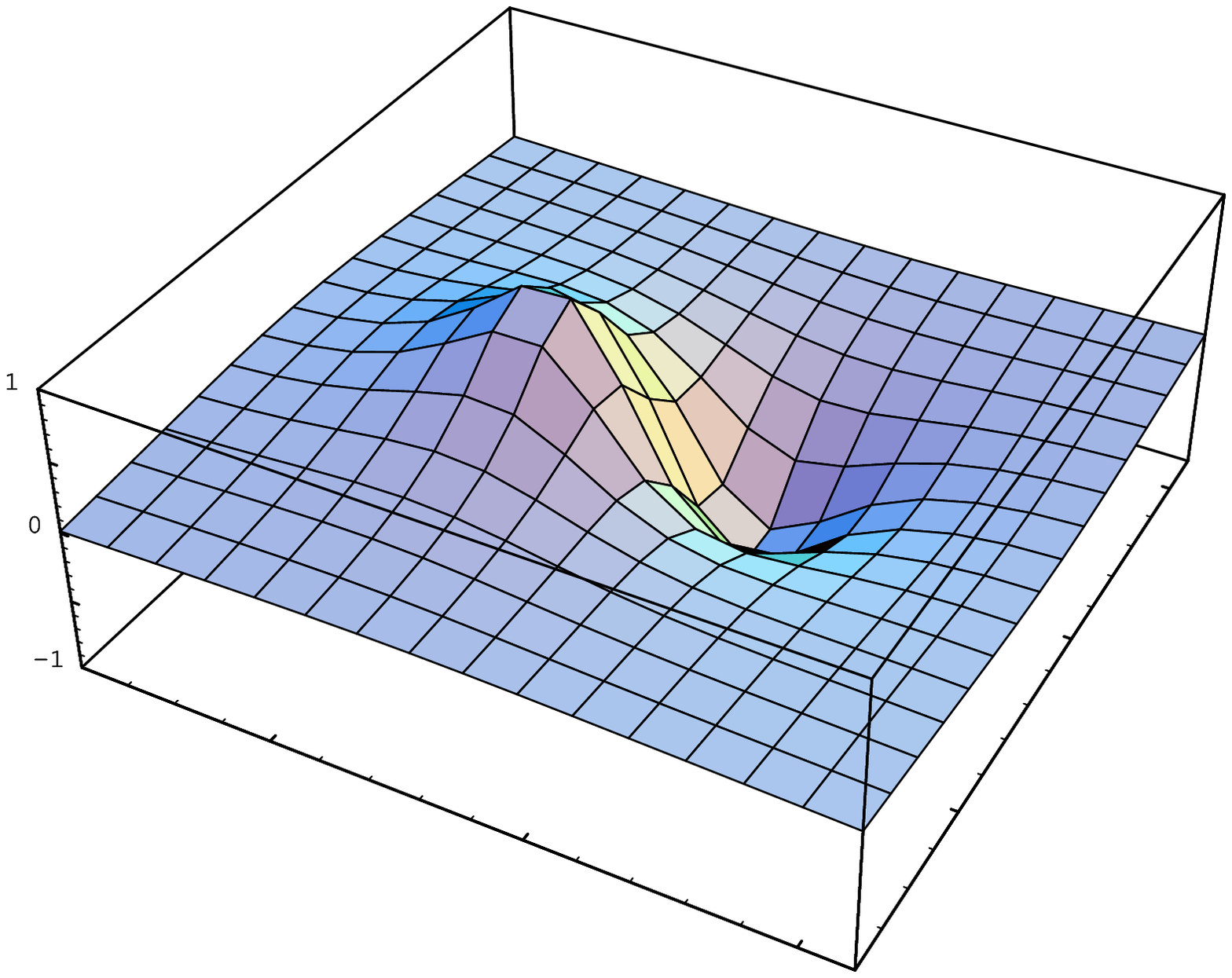}
\caption{Profile of the Polyakov loop $P_0(\vec x)$ for the calorons of fig.~5.
Left corresponds to $\vec k=(1,1,1)$ and $\omega=\quart$, right is for $\vec k
=(0,0,0)$ with $\omega=0.185$. Plotted is the plane $y=0$, for other details
see the caption of fig.~5.}
\end{figure}

For large $L/\beta$ the  difference between the  finite volume solutions 
with respect to the infinite volume calorons is mostly due to  the 
contribution of the Coulombic tails of the periodic
copies of the monopole constituents. That this depends on the nature of 
the twist is to be expected. For twist in time the charges change sign when 
shifting over a period of the torus. For twist in space there is no change 
in sign. This behaviour of the charges is correlated to the zeros of $A_0$ 
(which plays the role of the Higgs field) at the core of the constituent 
monopoles as illustrated by the behaviour of $P_0$ which is anti-periodic 
with time-twist and periodic with space-twist. 
It can be shown from the analytic solution (see the Appendix) that $P_0=1$ 
(corresponding to $A_0=0$) near one of the constituent centers, and $P_0=-1$ 
near the other (related to $A_0=0$ by a gauge transformation that is 
anti-periodic in time - the gauge transformation that changes $\omega$ 
to $\half-\omega$). This vanishing, i.e. $P_0^2=1$, of the Higgs field near
to the constituent monopole centers is reproduced by the lattice data as is 
illustrated in figure 7.

\section{Higher charge calorons}

In this section we discuss our finding for higher topological charge. Analytic
results in infinite volumes for higher charge calorons with non-trivial 
holonomy are not yet available. 
\begin{figure}[htb]
\vspace{10cm}
\includegraphics{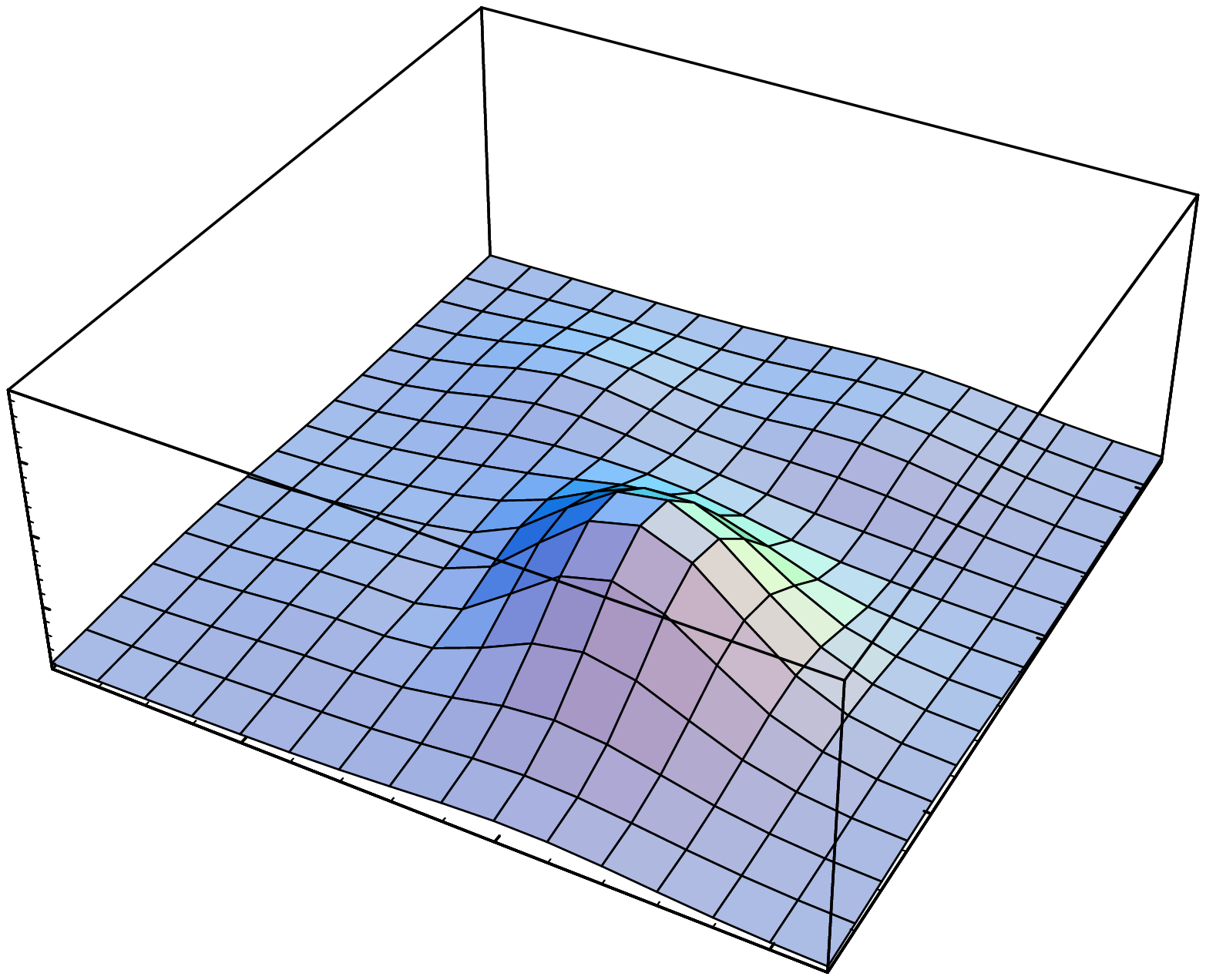}
\includegraphics{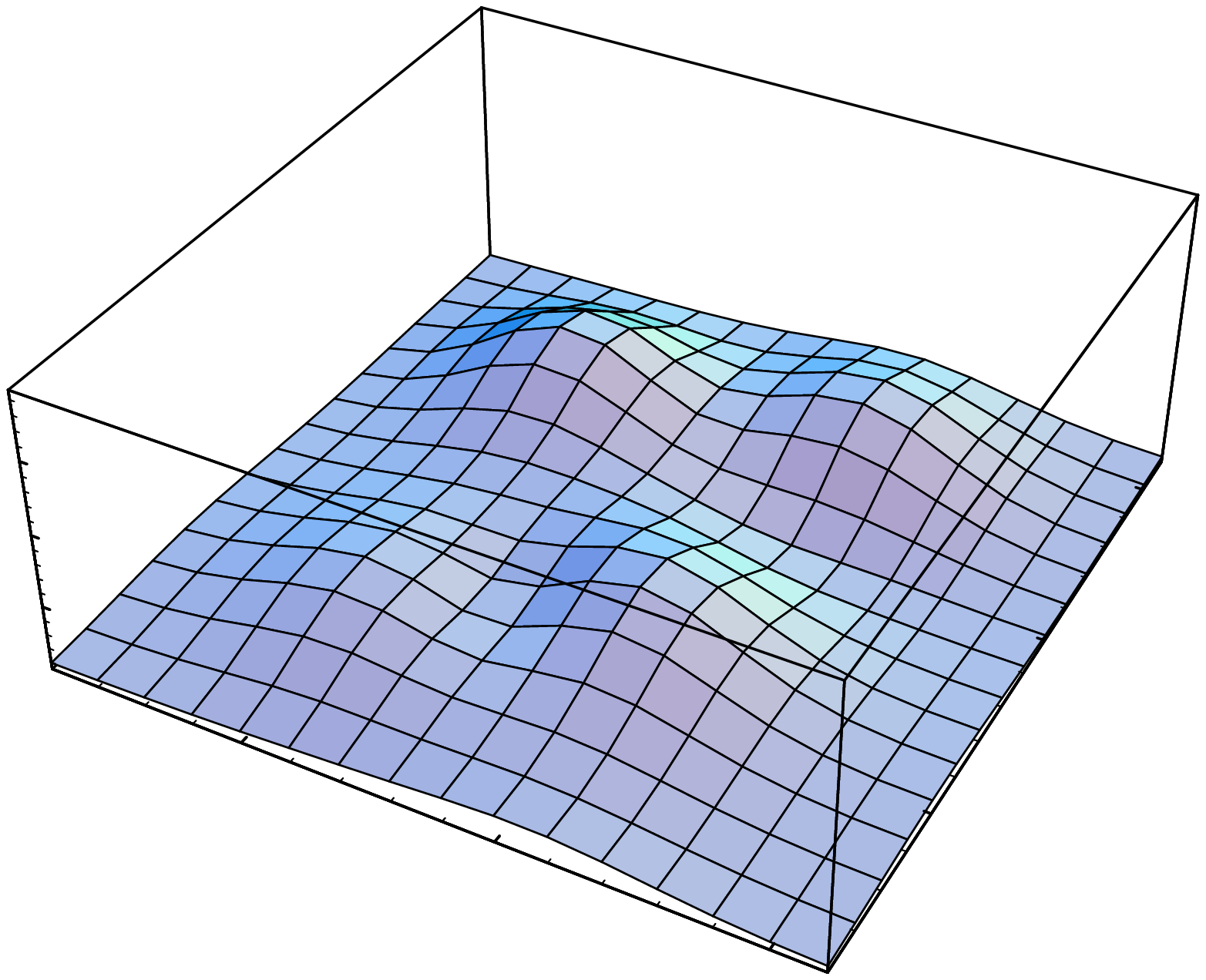}
\includegraphics{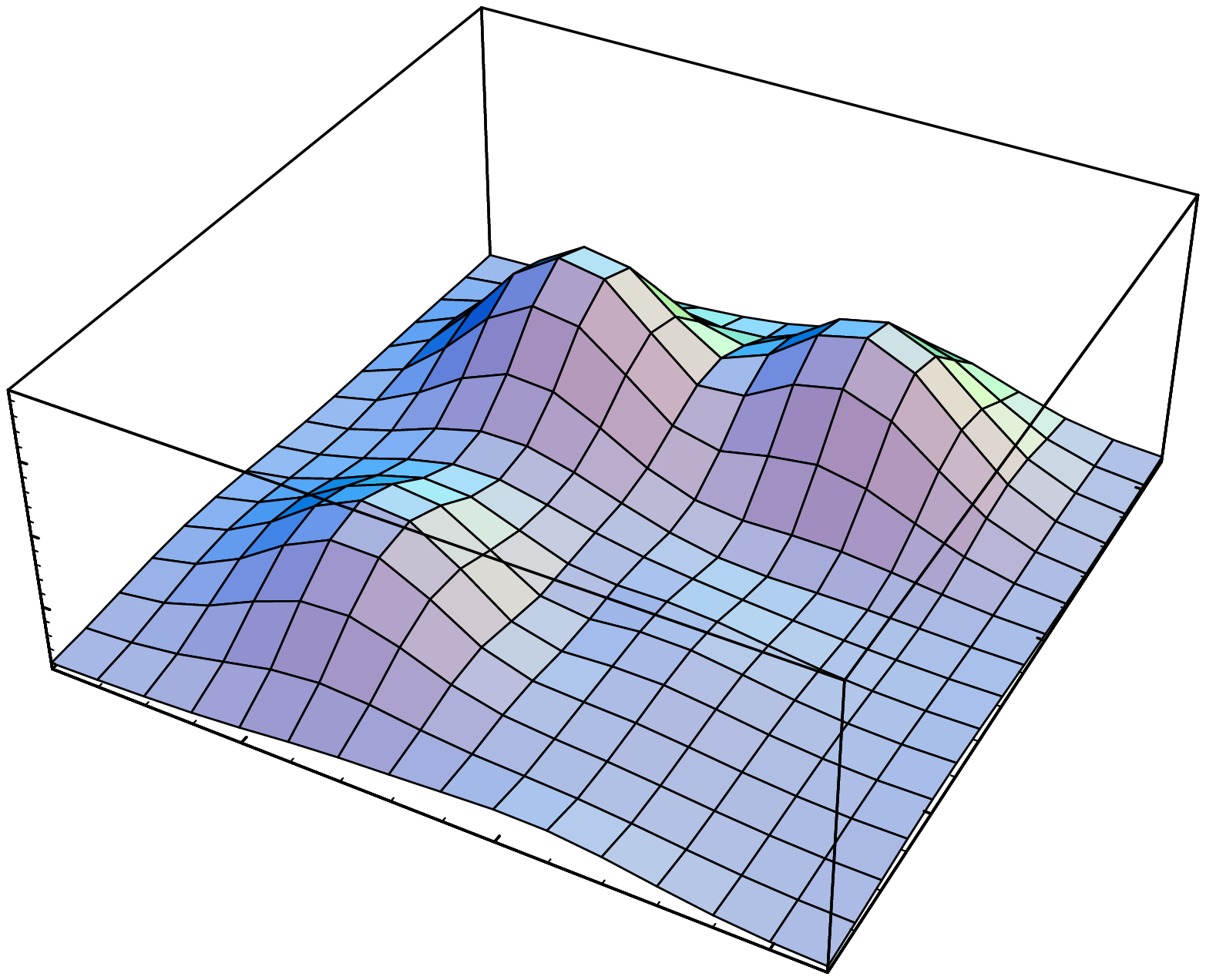}
\includegraphics{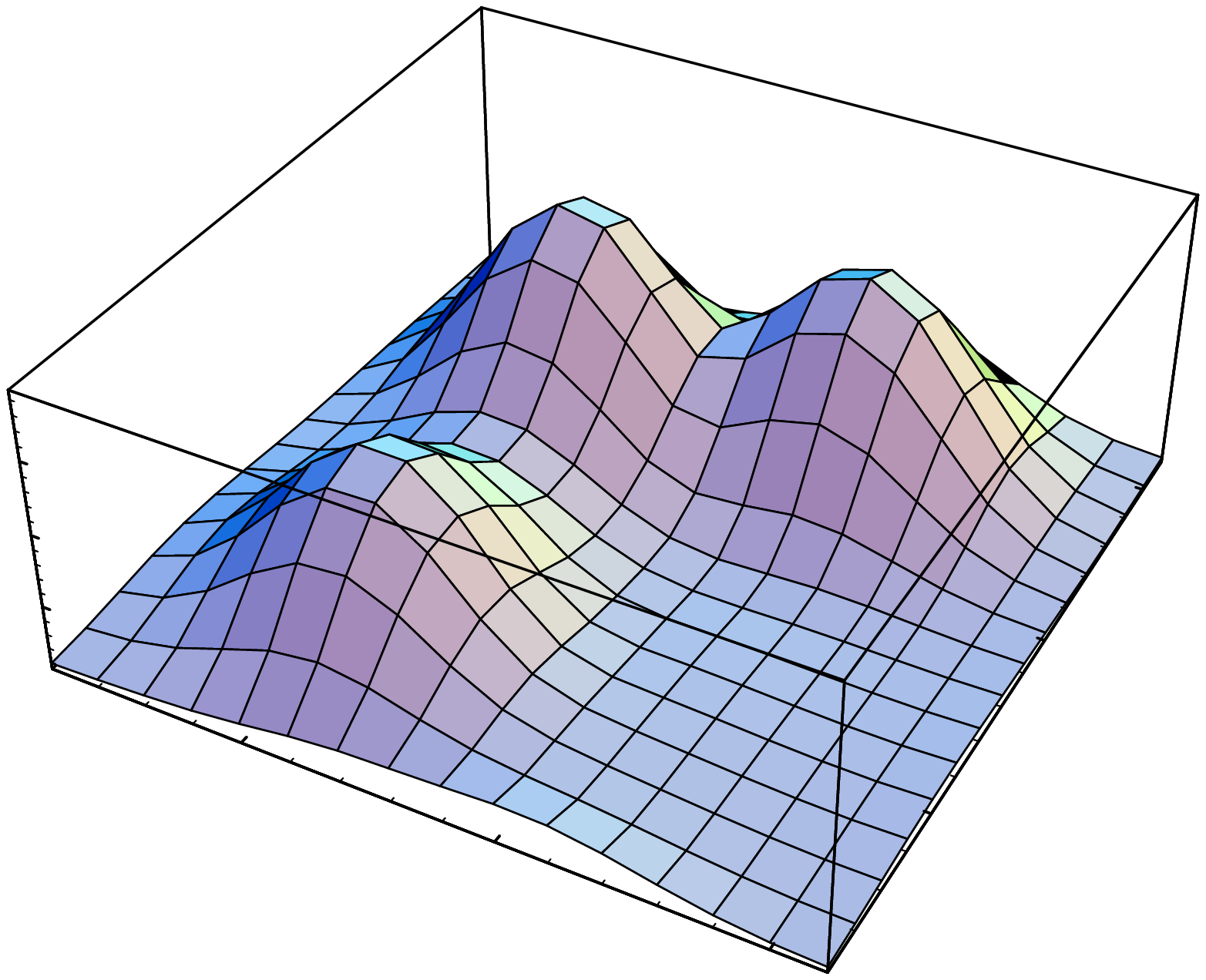}
\caption{Lattice action density profiles of a (static) charge 2 configuration
found with ordinary cooling on a $16^3\times 4$ lattice with $\vec k=(1,1,1)$, 
as a function of $x$ and $z$ for four consecutive slices in the $y$ direction.}
\end{figure}
Due to the local (lumpy) character of the 
caloron solutions one would expect that higher charge configurations can 
be obtained by ``gluing'' together lower charged solutions. Indeed for 
configurations on the torus, considering more than one period in any direction 
is a sure way of producing solutions with higher topological charge. On the 
basis of this it is to be expected that in the case of SU(2), for example,  
configurations would have $2Q$ action density lumps. 

Producing high charge 
configurations with our method is simple. It is sufficient to monitor the value
of the lattice action during cooling. Typically, this quantity shows plateaus 
at integer multiples of $8 \pi^2$. The cooling process can be interrupted at 
the desired value of the lattice action. We used ordinary ($\eps=1$) cooling; 
resulting configurations can subsequently be studied in more detail with 
other values of $\eps$. 

Figure 8 shows a configuration of charge 2, generated with ordinary cooling 
and twist $\vec k=(1,1,1)$. Indeed we find four lumps. We have been able to 
fit these to two $Q=1$, $\omega=\quart$, calorons by just adding the action 
densities together. Other charge 2 configurations have been obtained as well.
This includes a configuration with 3 lumps, one of which seems describable 
as a $Q=1$ object.

With similar techniques one can generate configurations with topological
charge higher than $2$. This process led us to study the whole cooling
histories  that go from randomly generated configurations to low action
ones. On lattices $N_s^3\times 4$, with $N_s=16$, 20 and 24, we computed
every 10 ($\eps=1$) cooling steps the total action $S$ of the configuration 
and used our peak-searching algorithm to locate action density maxima.
The information was recorded whenever the density of peaks, 
$N_{peak}/(N_s^3 \times 4)$, found by the algorithm was smaller than 
$50/(24^3 \times 4)$ (for higher densities the results are too sensitive 
to the details of the peak searching algorithm to be considered reliable). 
For all recorded data the quotient $S/(4 \pi^2 N_{peak})$ was found to 
lie between $0.8 $ and $2$, and peaked around $1$. This means that {\em on 
average} every peak is associated to an action of $4\pi^2$, a property shared 
with the exact $Q=1$ caloron solution with {\em non-trivial} holonomy. The same
follows for configurations that are aggregates of $Q=1$ calorons, which each 
have either one or (more often) two lumps (the constituent monopoles). 
Our result shows that this pattern extends to higher densities, where a 
detailed analysis of individual peaks is hard to do. Furthermore, the sign 
of the topological charge of these lumps is not always the same, thereby 
pushing the picture of a constituent monopole ensemble beyond the case of 
self-dual configurations. Our result resembles the findings of ref.~\cite{Ton2},
where a similar behaviour was reported for Monte Carlo generated configurations 
at zero temperature. In our case, we have the additional advantage of having 
an analytic control for $Q=1$ self-dual configurations. This allows us to 
conclude that the lumps correspond to constituent monopoles and hence, at 
least in this finite temperature case, not all lumps carry integer or 
half-integer topological charge. We hope these results will help to motivate 
other authors to investigate this point further.

\section{Discussion}
In this paper we have shown that $Q=1$ self-dual solutions can be obtained
profusely  on asymmetric lattices $L^3 \times \beta$ with $L \gg \beta$ by using
twisted boundary conditions. These configurations match quite well 
the analytic caloron solutions on $R^3\times S_1$~\cite{KSu2}. The main 
change induced by  the finite spatial volume is due to the contribution 
of the Coulombic tails of the
periodic mirrors of the caloron solutions. We have shown that with judicious
use of the twist values and of the parameter $\eps$ appearing in the
cooling method of ref.~\cite{UsCo}, one can produce caloron solutions with
different values of $\rho$ and $\omega$. In comparing to the continuum
expressions, the choice $\eps=0$ (improved cooling) reduces considerably the 
size of lattice corrections. 

In our analysis we have attempted to disentangle the finite size effects from
the lattice artifacts, by making use of the $\eps$ engineering. We have also
explored self-dual configurations with higher values of the topological charge. 
Our results show that these configurations look very much like ensembles of
$Q=1$ calorons with trivial or non-trivial holonomy. The conclusion, sustained
by our results, is that typically a configuration with topological charge
$Q$ has $2 Q$ lumps (constituent monopoles).  

Given their local nature and the non-perturbative nature of the QCD
(Yang-Mills) vacuum we vindicate that these configurations ought to play a role
in the dynamics of the theory. It is to be emphasised that for high charges, 
the existence of these solutions does not rely on the use of any particular 
boundary conditions (twisted or not). Twist however plays a role in stabilising
these solutions under cooling and this lies at the heart of the success of our 
method. This is most probably related to the fact that there are no exactly 
self-dual $Q=1$ solutions on the torus in the absence of twist~\cite{BrvB}. 
This does not happen for non-zero twist~\cite{MaTo,Braam}. Thus, lattice 
studies involving cooling methods could introduce distortions for low values 
of the topological charge~\cite{MaNuPh0}. We stress again that due to its 
simple implementation and zero computational overhead, the use of twisted 
boundary conditions is an ideal tool for non-perturbative investigations
of non-Abelian gauge theories and QCD.

\section*{Appendix}
Here we summarise the infinite volume analytic solutions for the $SU(N)$ 
calorons with non-trivial holonomy. After a constant gauge transformation, 
the holonomy $H$ is characterised by ($\sum_{m=1}^n\mu_m=0$)
\be
H=\exp[2\pi i{\rm diag}(\mu_1,\ldots,\mu_n)],\qquad
\mu_1<\ldots<\mu_n<\mu_{n\!+\!1}\equiv\mu_1+1.
\ee
Note that $\Tr(H)/N=\lim_{|\vec x|\rightarrow\infty} P_0(\vec x)$. Using the 
classical scale invariance to put $\beta=1$, one has
\be
s(x)=-\half\Tr F_{\mu\nu}^{\,2}(x)=-\half\partial_\mu^2\partial_\nu^2\log
\psi(x),\quad\psi(x)=\Psi(\vec x)-\cos(2\pi t),
\quad\Psi(\vec x)=\half\tr(A_n\cdots A_1),
\ee
where
\be
A_m\equiv\frac{1}{r_m}\left(\!\!\!\ba{cc}r_m\!\!&|\vec y_m\!\!-
\!\vec y_{m+1}|\\0\!\!&r_{m+1}\ea\!\!\!\right)\left(\!\!\!\ba{cc}c_m\!\!&
s_m\\s_m\!\!&c_m\ea\!\!\!\right).
\ee
Noting that $r_{n+1}\equiv r_1$ and $\vec y_{n+1}\equiv\vec y_1$ we defined
$r_m=|\vec x-\vec y_m|$, with $\vec y_m$ the position of the $m^{\rm th}$ 
constituent monopole, which can be assigned a mass $16\pi^2\nu_m$, where 
$\nu_m\equiv\mu_{m+1}-\mu_m$. Furthermore, $c_m\equiv\cosh(2\pi\nu_m r_m)$ 
and $s_m\equiv\sinh(2\pi\nu_m r_m)$.

Restricting to the gauge group of $SU(2)$, choosing $H=\exp(2\pi i\omega\tau_3)$
and defining $\pi\rho^2=|\vec y_2-\vec y_1|$, we can place the constituents at 
$\vec y_1=(0,0,\nu_2\pi\rho^2)$ and $\vec y_2=(0,0,-\nu_1\pi\rho^2)$ by a 
suitable combination of a constant gauge transformation, spatial rotation and 
translation. For this case the gauge field reads 
\be
A_\mu(x)=\frac{i}{2}\bar\eta^3_{\mu\nu}\tau_3\partial_\nu\log\phi(x)+
\frac{i}{2}\phi(x)\myre\left((\bar\eta^1_{\mu\nu}-i\bar\eta^2_{\mu\nu})
(\tau_1+i\tau_2)\partial_\nu\chi(x)\right),
\ee
where the anti-selfdual 't~Hooft tensor $\bar\eta$ is defined by 
$\bar\eta^i_{0j}=-\bar\eta^i_{j0}=\delta_{ij}$ and $\bar\eta^i_{jk}=
\varepsilon_{ijk}$ (with our conventions of $t=x_0$, $\varepsilon_{0123}=-1$) 
and $\tau_a$ are the Pauli matrices. Furthermore, $\phi^{-1}(x)=1-\frac{\pi
\rho^2}{\psi(x)}\left( \frac{s_1c_2}{r_1}+\frac{s_2c_1}{r_2}+\frac{\pi\rho^2
s_1s_2}{r_1r_2}\right)$ and $\chi(x)=\frac{\pi\rho^2}{\psi(x)}\left(e^{-2\pi 
it}\frac{s_1}{r_1}+\frac{s_2}{r_2}\right)e^{2\pi i\nu_1 t}$, with $\nu_1=
2\omega$ and $\nu_2=1-2\omega$. The solution is presented in the ``algebraic'' 
gauge, $A_\mu(t+1,\vec x)=\exp(2\pi i\omega\tau_3)A_\mu(t,\vec x)
\exp(-2\pi i\omega\tau_3)$. Since the radii $r_i$ are even functions of $x$ 
and $y$, derivatives in these two directions vanish on the $z$-axis. Hence,
along the line connecting the two constituents $A_0$ is Abelian, allowing
for a simple result for $P_0$ along this axis 
\be
P_0(z)=\cos(\pi\nu_1+\Phi(z)),\quad\Phi(z)=\half\int^1_0dt~\partial_z\log
\phi(t,z).
\ee
Since $\psi(x)$ and $\phi(x)$ are even functions of $r_i$ we may substitute
$r_i=z-z_i$ (with $z_1=\nu_2\pi\rho^2$ and $z_2=-\nu_1\pi\rho^2$) to find
$\phi(t,z)=(\Psi(z)-\cos(2\pi t))/(\cosh(2\pi z)-\cos(2\pi t))$, with
$\Psi(z)=\cosh(2\pi z)+\pi\rho^2\left(\frac{s_1c_2}{r_1}+\frac{s_2c_1}{r_2}+
\frac{\pi\rho^2s_1s_2}{r_1r_2}\right)>1$ a smooth function of $z$. The pole
of $\phi(x)$ at $x=0$ represents the usual gauge singularity. It leads to
a jump of $2\pi$ in $\Phi(z)$, to which the gauge invariant observable $P_0(z)$
is insensitive. The integration over time can be performed explicitly and one 
finds
\be
P_0(z)=-\cos\left[\nu_1\pi+\half\partial_z\acosh(\Psi(z))\right].
\ee
{}From this it is easily shown that each of the values $P_0(z)=\pm1$ is taken 
only once. Only for large $\rho$ one finds $P_0(z_1)=1$ and $P_0(z_2)=-1$. When 
associating the constituent monopole locations to the zeros of the Higgs field 
(i.e. to $P_0^2(\vec x)=1$), we find these are shifted {\em outward} from 
$\vec y_i$. This is illustrated in figure 9. For the cases we studied in this 
\begin{figure}[htb]
\vspace{5cm}
\includegraphics{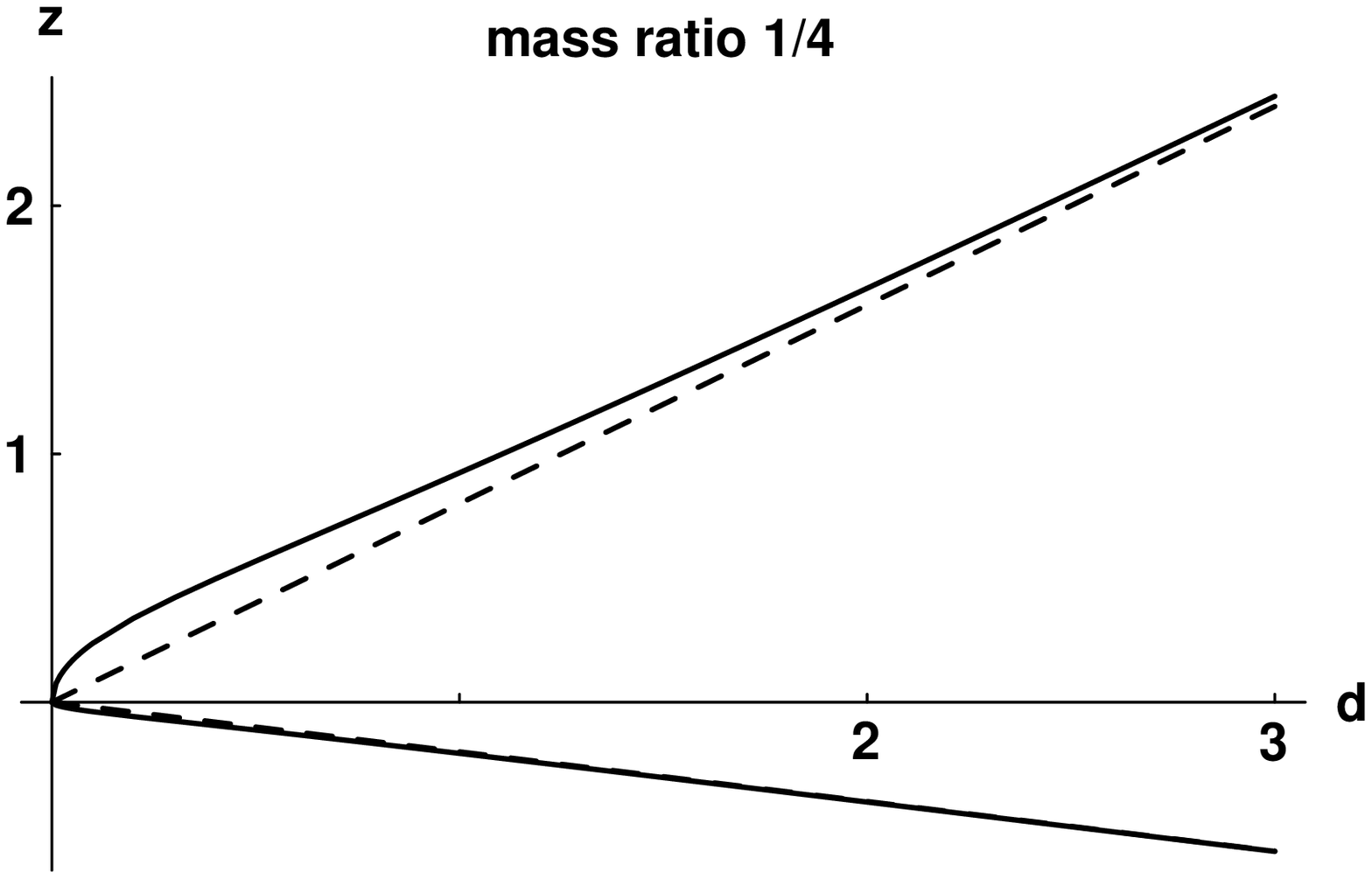}
\includegraphics{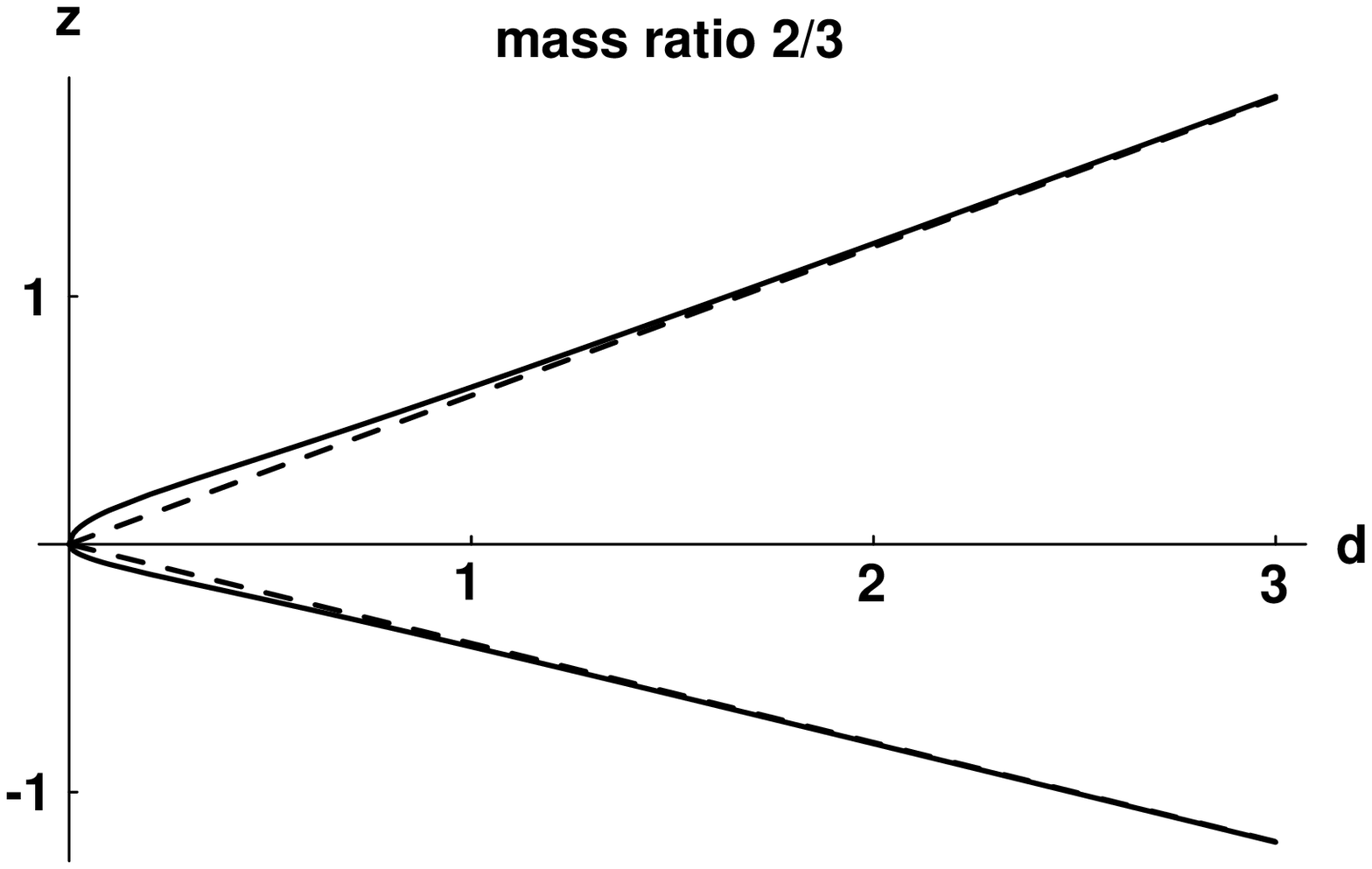}
\caption{Shift of the locations where $P_0^2(\vec x)=1$ as compared to the
location of the constituent monopole centers $\vec y_i$. Horizontally is
plotted the distance $d=\pi\rho^2$ between the constituents and vertically
the position of $z_1=(1-2\omega)d$ and $z_2=-2\omega d$ and the locations
where $P_0(z)=1$ ($z>0$) and $P_0(z)=-1$ ($z<0$). Left is for $\omega=0.1$
and right for $\omega=0.2$.}
\end{figure}
paper these shifts are small, but they tend to become large for the constituent
monopoles with a small mass ($\omega$ approaching either 0 or $\half$). We 
should also note that the maxima of the energy density (at $t=0$) are shifted 
{\em inward} due to overlap of the energy profiles of each constituent. 

The numerical evaluation of the action density $s(x)$ and of the Polyakov
loop $P_0(\vec x)$ are straightforward, but tedious. For the action density
it involves taking 4 derivatives, which is most conveniently achieved by
using the fact that $\Psi(\vec x)$ depends on $\vec x$ through the radii
$r_i$. The C-programmes written for this purpose are available~\cite{WWW}.

\section*{Acknowledgements}
We are grateful to Conor Houghton, Thomas Kraan and Carlos Pena for 
useful discussions. This work was supported in part by a grant from 
``Stichting Nationale Computer Faciliteiten (NCF)'' for use of the 
Cray Y-MP C90 at SARA. A. Gonzalez-Arroyo and A. Montero acknowledge 
finantial support by CICYT  under grant AEN97-1678. M. Garc\'{\i}a P\'erez
acknowledges finantial support by CICYT and warm hospitality at the Instituut
Lorentz while part of this work was developed.

\end{document}